\documentclass[11pt,a4paper,reprint,notitlepage,aps,prx,superscriptaddress]{revtex4-2}

\usepackage{amsmath,amssymb,amsfonts} % standard AMS packages

\usepackage[svgnames,dvipsnames]{xcolor}
\usepackage{graphicx}
\usepackage[colorlinks,
	linkcolor=red,
	citecolor=blue,
	urlcolor=red]{hyperref}

\usepackage{cleveref}

\usepackage{physics}

\def\omegam{\Omega_0}
\def\omegaeff{\Omega_{\mathrm{eff}}}
\def\gammam{\Gamma_0}

\def\gammaeff{\Gamma_{\mathrm{eff}}}

\def\neff{n_{\mathrm{eff}}}

\def\xsq{\langle \delta x^2 \rangle}
\def\psq{\langle \delta p^2 \rangle}

\def\nth{n_{\mathrm{th}}}

\def\neff{n_{\mathrm{eff}}}

\def\ntheff{n_{\mathrm{th,eff}}}
\def\nba{n_{\mathrm{ba}}}
\def\nimp{n_{\mathrm{imp}}}
\def\ncor{n_{\mathrm{cor}}}

\def\Ctot{C_\mathrm{tot}}
\def\Cimp{C_\mathrm{imp}}
\def\Ccor{C_\mathrm{cor}}

\def\Omegah{\Omega_{\mathrm{H}}}

\def\omegal{\Omega_{\mathrm{L}}}

\def\omegarp{\Omega_{\mathrm{rp}}}
\def\gammarp{\Gamma_{\mathrm{rp}}}

\def\gammafb{\Gamma_{\mathrm{fb}}}

\def\omegasql{\Omega_{\mathrm{SQL}}}

\def\Cqu{C_{Q}}
\def\gfb{g_{\mathrm{fb}}}

\def\thetaref{\theta_{\mathrm{eff}}}

\def\omegasqlzero{\Omega_{\mathrm{SQL},0}}
\def\Cqusql{C_{\mathrm{Q,SQL}}}

\renewcommand{\t}[1]{\text{{#1}}}
\newcommand{\avg}[1]{\left\langle {#1}\right\rangle}

\begin{document}

\title{Quantum theory of feedback cooling of an anelastic macro-mechanical oscillator}

\author{Kentaro Komori}
\email{komori.kentaro@jaxa.jp}
\affiliation{LIGO Laboratory, Massachusetts Institute of Technology, Cambridge, Massachusetts 02139, USA}
\affiliation{Institute of Space and Astronautical Science, Japan Aerospace Exploration Agency, Sagamihara, Kanagawa 252-5210, Japan}
\author{Dominika \v{D}urov\v{c}\'{i}kov\'{a}}
\affiliation{Department of Physics, Massachusetts Institute of Technology, Cambridge, Massachusetts 02139, USA}
\author{Vivishek Sudhir}
\email{vivishek@mit.edu}
\affiliation{LIGO Laboratory, Massachusetts Institute of Technology, Cambridge, Massachusetts 02139, USA}
\affiliation{Department of Mechanical Engineering, Massachusetts Institute of Technology, Cambridge, Massachusetts 02139, USA}

\date{\today}

\begin{abstract}
	Conventional techniques for laser cooling, by coherent scattering off of
	internal states or through an optical cavity mode, have so far proved inefficient on mechanical oscillators heavier
	than a few nanograms. That is because larger oscillators vibrate at frequencies much too small compared to the
	scattering rates achievable by their coupling to auxiliary modes. %, which makes laser sideband cooling inefficient.
	Decoherence mechanisms typically observed in heavy low frequency elastically suspended oscillators also differ markedly
	from what is assumed in conventional treatments of laser cooling. We show that for a low-frequency anelastic 
	oscillator forming the mechanically compliant end-mirror of a cavity, detuned optical readout, together with measurement-based
	feedback to stiffen and dampen it, can harness ponderomotively generated quantum correlations, to realize efficient
	cooling to the motional ground state.
	This will pave the way for experiments that call for milligram-scale mechanical oscillators prepared in pure
	motional states, for example, for tests of gravity's effect on massive quantum systems.  
\end{abstract}

% \begin{abstract}
% Mechanical oscillators with internal dissipation --- such as anelasticity --- ubiquitous in 
% Macroscopic superposition states can shed light on quantum physics in large gravitational fields. Cooling of a macroscopic mechanical mode is important for realizing the quantum state. Feedback cooling is required to reach quantum regime of the macroscopic oscillator. Here, we explore optimal feedback cooling limited by thermal, back-action, and quantum noise caused by the feedback, and derive quantum limit of the cooling.
% \end{abstract}

\maketitle

\section{Introduction}

The purity with which quantum states of tangibly massive objects can be prepared remains an 
open experimental challenge \cite{WhitSud21,Neu21,auriga08}. Although workers in the fields of 
atomic physics \cite{Died89,Monroe95,Perr98,Esch03,Maunz04,Booz06},
and more recently cavity optomechanics \cite{Chan2011,Teufel2011,Kauf12,Peterson2016,
Rossi2018,Delic20,Tebb20,MagAsp21,TebNov21}, 
have succeeded in addressing this challenge at sub-nanogram mass scales, 
objects with a significantly larger mass feature a qualitatively different behavior. The central pathology 
remains the same, namely decoherence, but the precise symptom is unique at large masses. 

Small-mass objects, elastically or electromagnetically bound, can be taken to be a mechanical oscillator that
is subject to a viscous damping force proportional to its velocity (called velocity damping).
For trapped atoms, this is due to the fact that there is little internal dissipation, and any external
dissipation arises predominantly from background gas collisions, which are naturally described through impulsive
momentum kicks; fluctuation-dissipation theorem then assigns a velocity-damped model for motional
decoherence. 
Levitated nano-mechanical systems, recently prepared in their motional ground state \cite{Tebb20,Delic20}, 
appear to be immune to internal dissipation, despite large internal temperatures \cite{Mill14}, apparently due 
to negligible coupling between internal modes and 
center-of-mass motion. Nano-mechanical objects are elastically bound so as to realize
radio-frequency mechanical oscillators; the effects of internal dissipation are largely masked at
such high frequencies \cite{Fed18}. The upshot is that all existing theoretical consideration of laser
cooling of mechanical oscillators implicitly assumes a velocity damped 
oscillator \cite{Manc98,Marq07,Wilson07,Genes2008}.

Large-mass objects have been isolated so as to be largely immune to external damping. 
To wit, gas damping (in the high Knudsen number --- ``low pressure'' --- regime) scales inversely
with the mass \cite{Christ66,Beres90,Cav10},
while suspension techniques have been developed (such as that employed in LIGO, or 
proposed schemes for levitation) that are not limited by external influence. 
% when an object of mass $m$ is placed
% in a gaseous environment of pressure $P$ sufficiently low such that the mean-free path of the 
% individual gas molecules is comparable to the typical size of the object 
% (the regime of the Knudsen number, $\t{Kn}\gtrsim 1$), 
% collisions with gas molecules --- gas damping --- dissipates energy at a 
% rate \cite{Christ66,Beres90,Cav10}, $\Gamma_\t{gas} \propto P/m$. 
Internal damping therefore dominates their decoherence. A most ubiquitous form of internal damping
in elastic oscillators is so called anelasticity \cite{Saul90,Fed18,Cripe2019}, for which the damping is not 
velocity-proportional, but is described by a frequency dependent ``structural damping'' 
rate, $\Gamma_0[\Omega]=(\Omega_0/Q)(\Omega_0/\Omega)$, where $\Omega_0$ is the resonance frequency, and
$Q$ is the (frequency-independent) quality factor.

% \begin{figure}[b!]
% 	\centering
% 	\includegraphics[width=\columnwidth]{fig_purity_v_mass_yzoom.pdf}
% 	\caption{\label{fig:survey}
% 		State of the art in preparing mechanical oscillators of various mass in pure quantum states. 
% 		Nanomechanical oscillators (top left corner) have been prepared in their motional ground states
% 		by cryogenic refrigeration \cite{Conn10,HongGro17,Chu17}, 
% 		laser cooling \cite{Chan2011,Teufel2011,Peterson2016}, 
% 		or active feedback cooling \cite{Rossi2018}. Other oscillators that have been realized in
% 		pure quantum states, such as levitated nanoparticles \cite{Tebb20,Delic20}, 
% 		optical phonons in crystals \cite{LeeWalm11,Velez19}, or
% 		single atoms \cite{Died89,Monroe95}, 
% 		are lighter by several orders of magnitude.
% 		The kg-scale test mass of Initial LIGO \cite{Abbott2009} and the ton-scale AURIGA bar 
% 		detector \cite{auriga08} were prepared in motional states that are few orders of magnitude impure.
% 	}
% \end{figure}

The decoherence rate of a structurally damped oscillator, when exposed to a thermal bath of mean
occupation $n_\t{th}[\Omega]\approx k_B T/\hbar \Omega$,
\begin{equation*}
	\Gamma_\t{th} [\Omega] = n_\t{th}[\Omega]\Gamma_0[\Omega] 
	\approx \frac{k_B T}{\hbar Q} \left( \frac{\Omega_0}{\Omega} \right)^2,
\end{equation*}
decreases quadratically with frequency, in marked contrast to a velocity damped oscillator (for which
the scaling is linear).
This can be harnessed by stiffening the oscillator
--- for example by radiation pressure forces from a cavity field \cite{BragKhal97,Corbitt2007,Ni12} ---
% for example by optically stiffening the oscillator using radiation pressure
% forces from a cavity field \cite{Corbitt2007,Ni12}. In this case, an oscillator mode
% can be established at a resonance frequency, 
so as to establish an oscillator mode at the frequency $\Omega_\t{eff} \gg \Omega_0$, whose thermal decoherence rate,
$\Gamma_\t{th} [\Omega_\t{eff}] = \Gamma_\t{th} [\Omega_0] (\Omega_0/\Omega_\t{eff})^2$, can be significantly lower than
that of the intrinsic mode (at frequency $\Omega_0$). 
%This low decoherence rate would facilitate laser cooling of the oscillator to its ground state if the additional 
However, this will be counteracted by additional 
decoherence from quantum fluctuations of the optical field used to produce the optical spring. 
The interplay of these two effects, given the scaling
of the decoherence rate for a structurally damped oscillator, call for a re-examination of the 
conventional theory of laser cooling \cite{Manc98,Marq07,Wilson07,Genes2008} as applied to 
macroscopic mechanical oscillators. As we will show, this naturally brings up the opportunity to
consider improvement of the cooling performance through back-action evasion.

% which decreases with frequency (here, $Q=\Gamma[\omega_0]/\omega_0$, is frequency independent). 
% In marked contrast to velocity damped oscillators, this means that when the oscillator is optically
% stiffened, from its natural resonance frequency $\omega_0$ to an effective one $\omega_\t{eff}$, the
% decoherence rate reduces by a factor $\omega_0/\omega_\t{eff}$, while the number of coherent oscillations
% it executes in one period, $N_\t{osc} = \omega_\t{eff}/\gamma[\omega_\t{eff}] \approx $

% Indeed the quest to realize a large mass object with low extraneous decoherence (and high purity) is precisely to 
% infer the existence of fundamental sources of external decoherence, 
% such as from gravity \cite{Diosi1984,Penrose1996,Bassi2017}.

In the following we study laser cooling of structurally damped and optically stiffened mechanical oscillators via
their coupling to an optical cavity field. 
%where the oscillator is structurally damped, and stiffened to 
Because typical macroscopic mechanical oscillators coupled to optical cavities
tend to be in the broadband cavity regime (i.e. mechanical frequency much lower than the cavity decay rate), 
cooling from cavity dynamical back-action is not practical to realize pure quantum states, so we consider
active feedback based on cavity-enhanced measurement of the oscillator position as the 
cooling mechanism \cite{Manc98,Genes2008,Wilson2015,Rossi2018}. 
In fact, significant optical stiffening, by blue-detuning the cavity mode it is coupled to, requires external 
feedback to stabilize the oscillator against parametric instabilities \cite{BragVya02}.
The natural rotation of the field quadratures due to cavity detuning, possibly enhanced by choice of
homodyne measurement angle to derive the error signal for feedback, gives rise to the possibility of enhancing
the performance of feedback cooling using quantum correlations developed intrinsically in the radiation-pressure
interaction \cite{HabHam16}.
% However, in contrast to prior treatments, we consider the mechanical oscillator to be structurally damped, 
% whose unique frequency response can be harnessed by stiffening the oscillator, either by 
% dynamical back-action from the cavity field, and/or, active external feedback. 

\section{Feedback cooling with active and detuned optical spring}
\label{sec:theory1}

\begin{figure}[t!]
	\centering
	\includegraphics[width=\columnwidth]{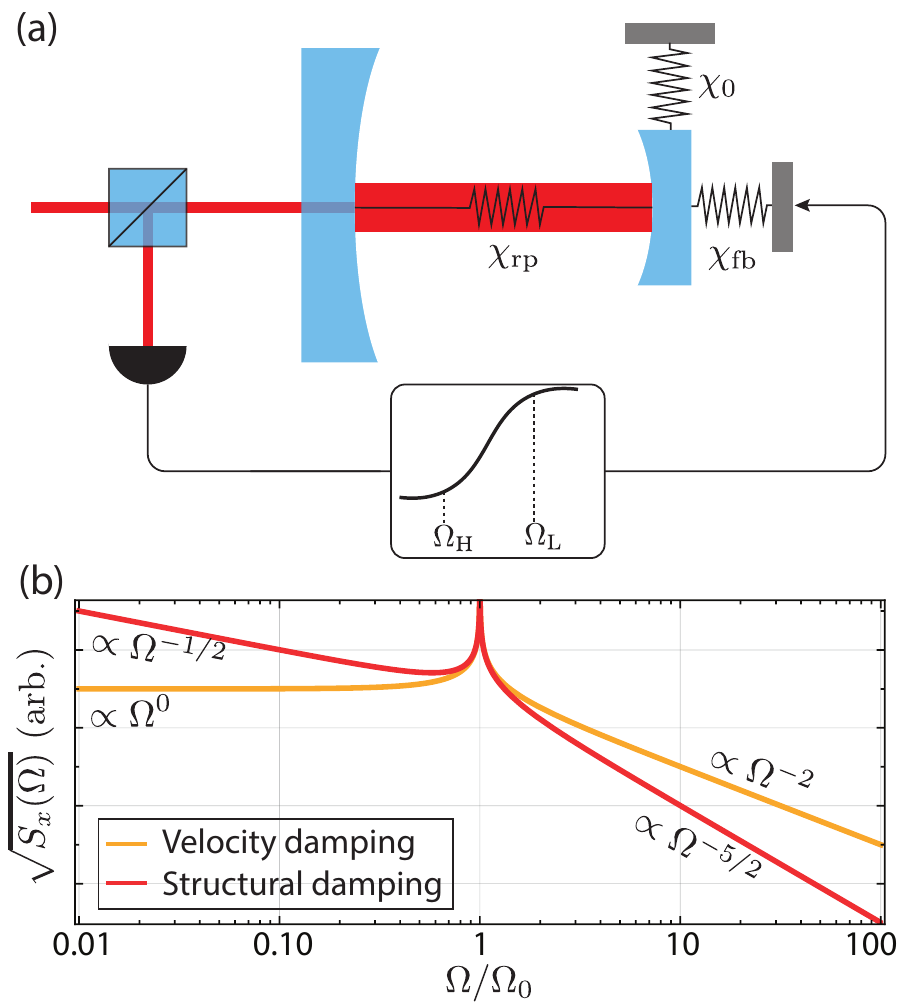}
	\caption{
	(a) A schematic picture of our model of feedback cooling. 
	A mechanical oscillator with intrinsic suscpetibility $\chi_0$ is trapped by two additional springs --- one induced by optical
	radiation pressure ($\chi_\t{rp}$) and another induced by measurement-based feedback ($\chi_\t{fb}$) 
	--- so that the effective resonant frequency is significantly increased. 
	The feedback filter is characterized by two cutoff frequencies $\Omega_\t{H,L}$ that define the mechanical
	mode of interest. 
	(b) Displacement fluctuations of a structurally-damped (red) and velocity-damped (yellow) oscillator, showing the
	stronger decrease of structural thermal noise at frequencies above resonance.
	}
	\label{fig:schematic}
\end{figure}

We consider here the following scenario (depicted in \cref{fig:schematic}a): 
a mechanical oscillator, with displacement fluctuations $\delta x$, forms
the end-mirror of an optical cavity, whose motion modulates the cavity frequency as $G\cdot \delta x$; the cavity is
probed by an ideal coherent field detuned from resonance by $\Delta$, and is otherwise lossless; the reflected light
is subjected to homodyne detection with a local oscillator whose phase differs from that of the cavity input
by $\theta$; the resulting photocurrent fluctuations are passed through a causal filter to synthesize a force --- 
the feedback force --- that impresses upon the oscillator. 
Despite the complexity of the scenario, the motion of the oscillator can be described by a simple
linear equation (in the frequency domain),
\begin{equation}\label{eq:x0}
	\chi_0^{-1}[\Omega] \delta x[\Omega] = \delta F_\t{th}[\Omega] + F_\t{rp}[\Omega] + F_\t{fb}[\Omega].
\end{equation}
It describes the intrinsic response, 
\begin{equation}
	\chi_0[\Omega] = [m(-\Omega^2 + \Omega_0^2 +i \Omega \Gamma_0[\Omega])]^{-1},
\end{equation}
of the oscillator --- with mass $m$, intrinsic resonance frequency $\Omega_0$, and damping $\Gamma_0[\Omega]$ --- 
to three forces.

The thermal force $\delta F_\t{th}$ is characterized by its (symmetrized double-sided) spectral density,
\begin{equation}
 	\bar{S}_{FF}^\t{th} [\Omega] 
 		%= 4\hbar (n_\t{th}[\omega]+\tfrac{1}{2}) \t{Im}\, \chi_0^{-1}[\omega]
 	 	= 2\hbar (n_\t{th}[\Omega]+\tfrac{1}{2}) m \Omega \Gamma_0[\Omega],
\end{equation} 
where, $n_\t{th}[\Omega] \approx kT/(\hbar \Omega)$, is the average thermal phonon occupation.
Note that structural thermal force decreases with frequency (i.e., $\bar{S}_{FF}^\t{th}[\Omega] \approx 2\hbar m\, 
\Omega \Gamma_\t{th}[\Omega] \propto 1/\Omega$, see \cref{fig:schematic}b).

The radiation pressure force $F_\t{rp}[\Omega]$ arises from an interaction between the oscillator displacement
and intracavity field ($a$) described by the interaction hamiltonian \cite{Aspelmeyer2014}, 
$H_\t{rp} = -\hbar G n x$, where $n=a^\dagger a$ is the intracavity photon number. 
In a linearized description, the radiation pressure force can be
expressed as the sum of two components \cite{BottStam12},
\begin{equation}\label{eq:Frp}
	F_\t{rp}[\Omega] = -\chi_\t{rp}^{-1}[\Omega] \delta x [\Omega] + \delta F_\t{rp}[\Omega].
\end{equation}
The first is a detuning-dependent force that is proportional to the oscillator position, and leads
to optical damping/anti-damping and spring shift, while the second is a quantum radiation pressure force
fluctuation due to intracavity photon number fluctuations. 
In the broadband cavity regime (i.e. where the cavity decay rate ($\kappa$) is much larger than the 
mechanical frequency)%, and with weak coupling (i.e. ) 
the former is described by a susceptibility of the form,
\begin{equation}\label{eq:chirp}
	\chi_\t{rp}^{-1} = m (\Omega_\t{rp}^2 + i \Omega \Gamma_\t{rp}),
\end{equation}
where, 
\begin{equation}
	\Omega_\t{rp}^2 \approx \Omega_\t{SQL}^2 \frac{\delta}{2(1+\delta^2)}, \qquad
	\Gamma_\t{rp} \approx -\frac{\Omega_\t{SQL}^2}{\kappa} \frac{\delta}{(1+\delta^2)^2},
\end{equation}
% \begin{align}
% 	\omega_\t{rp}^2 &\approx \omega_\t{SQL}^2 \frac{2\delta}{1+\delta^2} \\
% 	\gamma_\t{rp} &\approx \frac{\omega_\t{SQL}^2}{\kappa} \frac{4 \delta}{(1+\delta^2)^2},
% \end{align}
are the shifts in the oscillator frequency and damping rate due to the radiation pressure interaction.
Here, $\delta = \Delta/(\kappa/2)$ is the detuning normalized to the cavity's FWHM, $n_c$ is the mean
intracavity photon number, and we have defined, 
\begin{equation}\label{eq:omegaSQL}
	\Omega_\t{SQL}= \sqrt{ \frac{8\hbar G^2 n_c}{m \kappa}},	
\end{equation}
the frequency at which the free-mass standard quantum limit (SQL) is attained.

Note that in the theory of cavity optomechanics applied to high-frequency oscillators, the 
radiation-pressure-induced change in the oscillator frequency is typically small compared to the 
intrinsic resonance frequency (i.e. $\Omega_\t{rp} \ll  \Omega_0$). In that case, the characteristic
interaction frequency is the vacuum optomechanical coupling rate, $G\sqrt{n_c}\sqrt{\hbar/(2m\Omega_0)}$,
defined with respect to the zero-point motion of the intrinsic oscillator at frequency $\Omega_0$. 
In contrast, here we consider 
the scenario where the optical spring frequency can be much larger than the intrinsic frequency 
(i.e. $\Omega_\t{rp} \gtrsim \Omega_0$); and we are interested in the properties of the oscillator
mode established at the shifted frequency. 
Even when the shifted mechanical frequency is much different from its intrinsic frequency, the tradeoff between
measurement sensitivity and back-action force remains constrained by fundamental constants (explicated below).
Thus, in the terminology of imprecision and back-action noises (see below), the SQL frequency, implicitly defined 
by, $(m \omegasql)^2 \bar{S}_{xx}^\t{imp}[\omegasql] = \bar{S}_{FF}^\t{rp}[\omegasql]$, is a more convenient
measure of the interaction strength that is independent of the mechanical resonance frequency.
Note that this implicit definition clarifies the interpretation that it is the frequency at which the SQL is achieved.
When both the imprecision and back-action noises are white, for example for displacement readout using a 
broadband cavity, explicit expressions for these noises give the form in \cref{eq:omegaSQL}. The implicit
definition is however valid more generally.

The effect of the position-dependent term in the radiation pressure force (first term in \cref{eq:Frp}) 
is to change the effective response of the oscillator. Indeed, inserting the form of the radiation-pressure-modified
response [\cref{eq:chirp}] in \cref{eq:x0} and re-arranging terms shows that the radiation-pressure-modified response,
% produces,
% \begin{equation*}
%  (\chi_0^{-1} + \chi_\t{rp}^{-1}) \delta x = \delta F_\t{th} + \delta F_\t{rp} + F_\t{fb}.
% \end{equation*}
% Clearly, the new response, 
\begin{equation*}
	(\chi_0^{-1}+\chi_\t{rp}^{-1})^{-1} \approx [m(-\Omega^2 + (\Omega_0^2 +\Omega_\t{rp}^2) 
		+i \Omega(\Gamma_0 + \Gamma_\t{rp}))]^{-1},	
\end{equation*}
features a mechanical oscillator at a higher frequency $(\Omega_0^2 + \Omega_\t{rp}^2)^{1/2} \gg \Omega_0$ for
blue-detuned (i.e. $\delta > 0$) operation. Since the thermal force decreases with frequency, the displacement
fluctuations due to thermal noise at $\Omega_\t{rp}$ can be lower than that at the oscillator's instrinsic resonance
frequency $\Omega_0$.

Two effects however affect this conclusion. Firstly, quantum fluctuations in the intracavity photon number,
due to the blue-detuned light used to stiffen the oscillator, creates an additional radiation
pressure force fluctuation [see \cref{sec:ImpBA}], $\bar{S}_{FF}^\t{rp} = (\hbar G)^2 \bar{S}_{nn} 
\approx 4\hbar (\hbar G^2 n_c/\kappa)/(1+\delta^2)$, or equivalently,
\begin{equation}
\label{eq:SrpFF}
	\bar{S}_{FF}^\t{rp}[\omega] %\approx 4\hbar \frac{G^2 n_c}{\kappa}(1+\delta^2)^{-1} 
	= \hbar \frac{m \omegasql^2}{2(1 + \delta^2)}%\bigg \rvert_{\delta >0}
	= \hbar \frac{m \Omega_\t{rp}^2}{\delta}\bigg \rvert_{\delta >0} .
\end{equation}
Note that this quantum back-action noise increases quadratically with the stiffened oscillator frequency
(for fixed detuning and increasing laser power).
The second problem is that as the blue-detuned optical power is increased to realize a stiffer oscillator, 
the total damping rate can become negative (i.e. $\Gamma_0 + \Gamma_\t{rp} < 0$) and render the oscillator
unstable --- an example of radiation-pressure-induced parametric instability.

Both these problems --- increased quantum back-action with oscillator frequency, and parametric instability ---
can be controlled by applying a feedback force on the oscillator based on an estimate of its position.

The optical field used to pump the cavity --- the same one that when detuned produces the optical spring 
--- is modulated by the motion of the mechanical oscillator. Measuring the quadratures of the field
leaking out of the cavity, for example by homodyne detection, realizes a linear measurement of the 
mechanical oscillator's displacement. The homodyne photocurrent, appropriately normalied, produces 
a linear estimate of the position,
\begin{equation}
	\delta x_\t{obs}[\Omega] = \delta x [\Omega] + \delta x_\t{imp} [\Omega],
\end{equation}
contaminated by the displacement-equivalent imprecision noise $\delta x_\t{imp}$, due to shot noise fluctuations
of the field quadrature that is detected. Since the back-action force $\delta F_\t{rp}$ also arises from the
vacuum fluctuations of the same field, the imprecision and back-action satisfy two constraints (see \cref{sec:ImpBA}),
\begin{align}
	\bar{S}_{FF}^\t{rp}[\Omega] \bar{S}_{xx}^\t{imp}[\Omega] & 
		= \frac{\hbar^2}{4\eta }\csc^2 \theta_\t{eff} \label{eq:SF_Sx_1} \\
	\bar{S}_{Fx}^\t{rp,imp}[\Omega] 
		&= -\frac{\hbar}{2\sqrt{\eta}}\cot \theta_\t{eff} \label{eq:SF_Sx_2},
\end{align}
where
\begin{equation}
    \theta_\t{eff} = \theta -\tan^{-1}\delta
\end{equation}
is the effective quadrature angle of the reflected light that is measured (for example using a homodyne detector), and
$\eta$ is the detection efficiency.
The first expresses the essence of the uncertainty principle: the measurement imprecision and back-action force
are a mutual trade-off. The conventional measurement strategy --- for phase 
quadrature homodyne readout of the reflection ($\thetaref = \pi/2$) --- can realize a quantum-ideal measurement (i.e. $\bar{S}_{FF}^\t{rp} \bar{S}_{xx}^\t{imp} = \hbar^2/4$) if the detection efficiency is unity.
The second expression relays the fact that the back-action force and imprecision noise can be correlated --- 
but only for finite detuning and/or homodyne readout of non-phase quadratures --- due to the fact that 
traces of the same optical field fluctuations that produce the back-action force manifest also in the imprecision
noise. When these are anti-correlated (i.e. $\bar{S}_{Fx}^\t{rp,imp} < 0$), the detected field quadrature can 
be squeezed, and (some of) the back-action of the measurement avoided.
\Cref{eq:SF_Sx_1,eq:SF_Sx_2} exhaustively characterize the constraints on the measurement due to quantum 
mechanics at the level of spectral densities; in fact they verify the generalized
uncertainty principle, $\bar{S}_{FF}^\t{rp} \bar{S}_{xx}^\t{imp} - \vert\bar{S}_{Fx}^\t{rp,imp}\vert^2 
= \hbar^2/4$ \cite{brag,clerk10}.

Finally, a feedback force can be applied on the mechanical oscillator, based on such a measurement, i.e.,
\begin{equation}\label{eq:Ffb}
	F_\t{fb}[\Omega] = -\chi_\t{fb}[\Omega] \delta x_\t{obs}[\Omega],
\end{equation}
where $\chi_\t{fb}[\Omega]$ is a causal function chosen to produce the desired modification of the oscillator's
effective response. 
(In principle there could be an additional force noise associated with the feedback force ---
for example from the actuator in the feedback path, or technical noises in the photocurrent inside the passband
of $\chi_\t{fb}^{-1}$ ---
but this can always be made negligible as long as a sufficiently high-power quantum-noise-limited local oscillator
is used in the homodyne detector. In this case, the homodyne detector acts as a high-gain phase-sensitive
amplifier, and so the quantum noise of the optical field, assimilated into $\delta x _\t{imp}$, is the only
relevant noise.)
Inserting \cref{eq:Frp,eq:Ffb} in \cref{eq:x0} produces the equation of motion modified by radiation pressure
and feedback:
\begin{equation}
	\chi_\t{eff}^{-1}[\Omega] \delta x[\Omega] = \delta F_\t{th}[\Omega] + \delta F_\t{rp}[\Omega]
		- \chi_\t{fb}^{-1} \delta x_\t{imp}[\Omega],
\end{equation}
where $\chi_\t{eff}[\Omega]$ is the effective response given by,
\begin{equation}
\label{eq:chieffdef}
	\chi_\t{eff}^{-1} = \chi_0^{-1} + \chi_\t{rp}^{-1} + \chi_\t{fb}^{-1}.
\end{equation}
In order to affect active spring stiffening and cooling, the feedback susceptibility needs to 
approximate the form,
\begin{equation}\label{eq:chifb}
	\chi_\t{fb}^{-1} = m(\Omega_\t{fb}^2 + i\Omega \Gamma_\t{fb}),
\end{equation}
around the oscillator's stiffened frequency; here $\Omega_\t{fb},\Gamma_\t{fb} > 0$.
This form is comparable to the radiation-pressure-induced susceptibility in \cref{eq:chirp}.
The effective susceptibility then takes the form,
\begin{equation}
	\chi_\t{eff}^{-1} = m(-\Omega^2 + \Omega_\t{eff}^2 +i \Omega \Gamma_\t{eff}),
\end{equation}
which is the response of an oscillator at the shifted frequency,
$\Omega_\t{eff} = \Omega_0 + \Omega_\t{rp} + \Omega_\t{fb}$, with a modified damping rate,
$\Gamma_\t{eff}[\Omega] = \Gamma_0[\Omega] + \Gamma_\t{rp} + \Gamma_\t{fb}$.

The displacement spectrum of the oscillator so realized takes the form,
\begin{equation}
\label{eq:dispspec_all}
\begin{split}
	\bar{S}_{xx}[\Omega] =& \abs{\chi_\t{eff}[\Omega]}^2  \Big( \bar{S}_{FF}^\t{th} [\Omega] 
		+ \bar{S}_{FF}^\t{rp} [\Omega]
		+ \abs{\chi_\t{fb}[\Omega]}^{-2} \bar{S}_{xx}^\t{imp}[\Omega] \\
		& + 2\t{Re}\left(  \chi_\t{fb}^{-1}[-\Omega] \bar{S}_{Fx}^\t{rp,imp}[\Omega] \right) \Big).
\end{split}
\end{equation}
Here the first line represents the physical motion of the oscillator due to the thermal, radiation pressure 
back-action, and `feedback back-action' forces; the latter is due to imprecision noise fedback as a force through 
the filter $\chi_\t{fb}^{-1}$. The second term is due to imprecision-back-action correlations arising from 
detuning of the cavity from resonance, or detuning of the homodyne detector from phase quadrature.

When the objective is to cool the mechanical oscillator, a convenient figure of merit is the
average phonon number, $n_\t{eff}$, defined through the average energy,
\begin{equation*}
	\frac{\avg{\delta p^2}}{2m} + \frac{m\Omega_\t{eff}^2 \avg{\delta x^2}}{2} \equiv 
	\hbar \Omega_\t{eff} \left(  n_\t{eff}+\frac{1}{2} \right);
\end{equation*}
here $\delta p$ is the fluctuation in the momentum of the oscillator, which is unobserved. However, it can be estimated
from the observed displacement as, $\delta p[\Omega] = -i m \Omega\, \delta x[\Omega]$, so that the required
variances $\avg{\delta x^2}, \avg{\delta p^2}$ can be inferred from the spectral density $\bar{S}_{xx}$ alone as,
% $\avg{\delta x^2} = \int \bar{S}_{xx}[\omega]\, \frac{\dd \omega}{2\pi}$, and, 
% $\avg{\delta p^2} = \int (m \omega)^2 \bar{S}_{xx}[\omega]\, \frac{\dd \omega}{2 \pi}$.
\begin{gather}\label{eq:defxp}
	%\label{eq:defx}
	\xsq = \int^{\infty}_{-\infty} \frac{d\Omega}{2\pi} \bar{S}_{xx}[\Omega],\,\,
	%\label{eq:defp}
	\psq = \int^{\infty}_{-\infty} \frac{d\Omega}{2\pi} (m \Omega)^2 \bar{S}_{xx}[\Omega].
\end{gather}
Mathematically carrying out this program to estimate the phonon number for a structurally damped oscillator
that is controlled with the feedback filter in \cref{eq:chifb} turns out to be impossible. That is for two reasons:
\begin{enumerate}
	\item At low frequencies, even without feedback, the variance in the displacement of a structurally damped 
	oscillator is formally infinite \cite{Saul90}. The physical reason is that anelastic damping, just like
	any physical process dominated by $1/f$ noise, is due to non-equilibrium processes at slower and slower
	time-scales \cite{Pre50,mandelbrot_fractional_1968,nelkin_deviation_1981} which precludes thermal equilibrium.
	\item At high frequencies, feedback of imprecision noise as a force noise leads to
	a formally infinite momentum \cite{VitTom03}. This can be seen as follows: when \cref{eq:dispspec_all} is used
	to estimate the momentum variance as the integral of $(m \Omega)^2 S_{xx}$, the term in the integrand
	propotional to the imprecision noise, 
	$\Omega^2 \abs{\chi_\t{eff}}^2 \abs{\chi_\t{fb}}^{-2} \bar{S}_{xx}^\t{imp}$, is a constant at high
	frequencies, since $\abs{\chi_\t{eff}[\Omega \gg \Omega_\t{eff}]}^2 \sim \Omega^{-4}$, while,
	$\abs{\chi_\t{fb}[\Omega \gg \Omega_\t{eff}]}^{-2} \sim \Omega^2$, and (at best) $\bar{S}_{xx}^\t{imp}$ is
	frequency independent.
\end{enumerate}
In other words, a structurally damped oscillator does not strictly satisfy the equipartition principle; naive
feedback compounds the problem. 

In practice, all experiments have a finite bandwidth and observation time which
regulates the singularities at high and low frequencies respectively. In particular, for a large spring
($\Omega_\t{eff} \gg \Omega_0$) the effect of structural damping can be well approximated by taking the damping
rate to be constant around resonance, i.e. $\Gamma_0[\Omega] \approx \Gamma_0[\Omega_\t{eff}] 
= \Omega_0^2/(Q \Omega_\t{eff}) $. To regulate the problem with the momentum variance, we modify the feedback filter
to the form,
\begin{equation}\label{eq:chifb_new}
	\chi_\t{fb}^{-1}[\Omega] = m\Omega_0^2 \frac{1 + i \Omega/\Omegah}{1 + i \Omega/\omegal} \gfb,
\end{equation}
where $\Omegah,\omegal$ are high- and low-pass frequencies between which feedback is active ($\omegal > \Omegah$), 
%such that $\Omegah \gg \omega_\t{eff}\gg \omegal$ ($\omegal$) is a high (low) pass cutoff frequency, 
%$K_0 = m\Omega_0^2$ is the mechanical spring constant, 
and $g_\t{fb} > 0$ is the dimensionless gain. In this case,
unlike the naive filter in \cref{eq:chifb}, we have that $\abs{\chi_\t{fb}[\Omega \gg \Omega_\t{eff}]}^{-2} 
\sim m\Omega_0^2 \gfb^2$, so that, $\Omega^2 \abs{\chi_\t{eff}}^2 \abs{\chi_\t{fb}}^{-2} 
\bar{S}_{xx}^\t{imp} \sim \Omega^{-2}$, which regulates the 
high-frequency divergence of the momentum integral.
However, in order to realize a spring and damping, the filter in \cref{eq:chifb_new} must conform to the form
in \cref{eq:chifb} at some frequencies; indeed we have,
\begin{equation*}
	\chi_\t{fb}^{-1}[\Omega \ll \omegal] \approx 
	m\Omega_0^2 \left( 1 + i \frac{\Omega}{\Omegah} \right) \gfb.
\end{equation*}
Comparing this with \cref{eq:chifb} implies that the feedback damping is $\Gamma_\t{fb} = \gfb \Omega_0^2 /\Omegah$, and
the spring shift is, $\Omega_\t{fb} = \sqrt{g_\t{fb}\Omega_0^2} = \sqrt{\Omegah \Gamma_\t{fb}}$.

An additional complication of this choice of the feedback filter is that it need not render the system unconditionally 
stable in the presence of radiation pressure back-action. A simple Routh-Hurwitz analysis of the effective
suscpetibility $\chi_\t{eff}$ shows that the system is stable if, $\gfb > -\Gamma_\t{rp} \omegal \Omegah/[\omegam^2 (\omegal - \Omegah)]$. We assume that sufficient feedback damping can be realized to satisfy
this condition.

With these issues addressed, the oscillator's mean phonon number can be computed from the displacement spectral
density.  
The result can be expressed in closed form (see \cref{sec:AppA}):
\begin{widetext}
\begin{equation}\label{eq:neff}
\begin{split}
	2n_\t{eff}+1 = 
	\biggl[ &\frac{2\omegaeff^2 + (\omegal - \Omegah) \gammaeff + 2\omegal^2}{\omegal^2} 
	\left( \ntheff + \nba + \frac{1}{2} \right) \frac{\gammam[\omegaeff]}{\gammaeff}\\
	& + \frac{2\omegaeff^2 + (\omegal - \Omegah) \gammaeff + 2\Omegah^2}{\omegaeff^2} \nimp 
		\frac{\gammaeff}{\gammam[\omegaeff]} \\
	& - \frac{2\omegaeff^2 + (\omegal -\Omegah) \gammaeff + 2\omegal \Omegah}{\omegal \omegaeff} n_\t{cor} \biggr] 
		\left( 1 - \frac{\Omegah}{\omegal} \right)^{-1}.
	% \biggl[ \left( 1 + \Cqu^{-1} \right) \frac{2\omegaeff^2 + (\omegal - \Omegah) 
	% \gammaeff + 2\omegal^2}{\omegal^2 (1+\delta^2)} \frac{\omegasql^2}{\omegaeff \gammaeff}
	% &+ \frac{2\omegaeff^2 + (\omegal - \Omegah) \gammaeff + 2\Omegah^2}{16 \eta \omegasql^2} 
	% \left( 1 + \delta^2 \right) \csc^2 \thetaref \frac{\gammaeff}{\omegaeff} \\
	% & - \frac{2\omegaeff^2 + (\omegal -\Omegah) \gammaeff + 2\omegal \Omegah}{2 \sqrt{\eta} \omegal \omegaeff} 
	% \cot \thetaref \biggr] \left( 1 - \frac{\Omegah}{\omegal} \right)^{-1}.
	% 2n_\t{eff}+1 = 
	% \biggl[ \frac{2\omegaeff^2 + (1 - \omegah) \Omega_0^2 \gfb + 2\omegal^2}{\omegal} \frac{\gammam}{\Omega_0^2} \frac{\left( \ntheff + \nba \right)}{\gfb}
	% & + \frac{2\omegaeff^2 + (1 - \omegah) \Omega_0^2 \gfb + 2 \omegah^2 \omegal^2}{\omegal} \frac{\Omega_0^2}{\omegaeff^2 \gammam} \nimp \gfb \\
	% & - \frac{2\omegaeff^2 + (1 -\omegah) \Omega_0^2 \gfb + 2\omegah \omegal^2}{\omegal \omegaeff} n_\t{cor} \biggr] 
	% \left( 1 - \omegah \right)^{-1}.
\end{split}
\end{equation}
\end{widetext}
Here, $\ntheff = n_\t{th}[\Omega_\t{eff}] = n_\t{th}[\Omega_0](\Omega_0/\Omega_\t{eff})$, 
is the average phonon occupation of the stiffened oscillator,
$n_\t{ba} = \ntheff \cdot \bar{S}_{FF}^\t{rp}/\bar{S}_{FF}^\t{th}[\omegaeff]$ is the average phonon occupation
due to quantum back-action, $n_\t{imp}$ is the phonon-equivalent imprecision noise defined through the
uncertainty relation [\cref{eq:SF_Sx_1}],
$n_\t{imp} n_\t{ba} = (16\eta)^{-1}\csc^2 \theta_\t{eff}$, and $n_\t{cor} = (2\sqrt{\eta})^{-1} \cot \theta_\t{eff}$
is the phonon-equivalent correlation between imprecision and back-action.

\section{Discussion}

The first and second terms in \cref{eq:neff} denote the feedback-suppression of the total energy
of the stiffened oscillator ($\propto \ntheff + \nba$) and the heating due to feedback-injection of imprecision
noise, respectively. The third term, negative in contribution, is the effect of back-action cancellation originating from
imprecision-back-action correlations developed through the radiation pressure interaction \cite{SudKip17,KamReg17,MasSch19}.
Such quantum correlations can be harnessed when feedback is predicated on readout of the outgoing field's quadrature
that is away from phase quadrature (as shown in Ref. \cite{HabHam16} for feedback damping with resonant cavity readout
for a velocity-damped oscillator). 

\subsection{Conventional case: feedback with resonant phase-quadrature readout}

Before delving into further discussion, note first that the practice of estimating the phonon occupation by assuming
the equipartition principle, i.e. taking $2n_\t{eff}+1 = (2m \omegaeff/\hbar)\avg{\delta x^2}$, is equivalent to taking
the low-pass cutoff to be $\omegal \rightarrow \infty$; in this case, \cref{eq:neff} reduces to,
\begin{align*}
	\neff + \frac{1}{2} \approx & \left( \ntheff + \nba + \frac{1}{2} \right) \frac{\gammam[\omegaeff]}{\gammaeff} \\
	&+ \left( 1 + \frac{\Omegah^2}{\omegaeff^2} \right) \nimp \frac{\gammaeff}{\gammam[\omegaeff]}
	- \frac{\Omegah}{\omegaeff} n_\t{cor}.
\end{align*}
For phase measurement at zero-detuning (i.e. $\theta_\t{eff} = \pi/2, \delta =0$), the effective resonance frequency is
$\omegaeff^2 \approx \omegal \gammaeff$, so that the above expression can be cast as,
\begin{equation}\label{eq:neff:phase}
\begin{split}
	\neff + \frac{1}{2} \approx & \left[ \ntheff + \nba + \frac{1}{2} 
		+ \left( \frac{\omegaeff}{\gammam[\omegaeff]} \right)^2 \nimp \right] \frac{\gammam[\omegaeff]}{\gammaeff}\\ 
		&+ \nimp \frac{\gammaeff}{\gammam[\omegaeff]},
\end{split}
\end{equation}
consistent with the experiments on feedback cooling of a structurally damped actively stiffened 
oscillator near its ground state \cite{WhitSud21}. 

The case of no active spring corresponds to setting $\omegaeff =0$ (because
we have assumed that $\omegaeff \gg \omegam$ in the above equation). With resonant readout, there is no additional
source of spring stiffening either.
In this case, the above equation reduces to,
$(n_\t{eff}+\tfrac{1}{2})\gammaeff \approx (n_\t{th}+n_\t{ba} +\tfrac{1}{2})\Gamma_0 + n_\t{imp}\gammaeff$, 
which can be interpreted as a detailed balance relation describing a velocity-damped oscillator simultaneously 
coupled to its thermal and back-action baths at rate $\gammam$, and via feedback to the bath due to measurement 
imprecision at rate $\gammaeff$. 
Optimizing over the damping rate shows that, $n_\t{eff}+\tfrac{1}{2}\gtrsim 2\sqrt{(n_\t{th}+n_\t{ba}+\tfrac{1}{2})n_\t{imp}}$;
using the uncertainty relation, $n_\t{imp}n_\t{ba}\geq \tfrac{1}{16}$ further gives, $n_\t{eff}+\tfrac{1}{2}\gtrsim 
2 \sqrt{(n_\t{th}+\tfrac{1}{2})n_\t{imp}+\tfrac{1}{16}}$. Thus, to realize $n_\t{eff} < 1$ requires that 
$n_\t{imp}< 1/(2n_\t{th}+1)$, which is the well-understood requirement on the measurement sensitivity to feedback cool a
velocity-damped oscillator to its motional ground state \cite{Wilson2015,Rossi2018}.

In marked contrast, for a structurally damped oscillator that is
actively stiffened, the apparent initial occupation (i.e. before feedback damping has commenced) in
\cref{eq:neff:phase}, 
$n_\t{th,eff}+n_\t{ba}+n_\t{imp}(\omegaeff/\gammam[\omegaeff])^2$, has a thermal component (first term),
$\ntheff = \nth[\omegam](\omegam/\omegaeff)$, that decreases with increasing spring frequency --- a form of
thermal noise dilution \cite{Corbitt2007,Ni12},
and an additional term (third term), $n_\t{imp}(\omegaeff/\gammam[\omegaeff])^2 = n_\t{imp}(\omegaeff^2/\gammam \omegam)^2$, 
that increases with the spring frequency --- 
a form of feedback-back-action arising from imprecision noise fedback as a force noise in realizing the active spring.
The opposing scaling of these two effects with the spring frequency, with the former scaling as $\omegaeff^{-1}$ and the latter
as $\omegaeff^4$, implies an optimal value of the spring frequency beyond which the dilution of thermal noise
is nullified by increase in feedback-back-action from the spring. For a given measurement imprecision, which is independent from the effective frequency for the structurally damped oscillator
\begin{equation}
    \nimp = \frac{1}{4\eta Q_0} \left( \frac{\omegam}{\omegasql} \right)^2,
\end{equation}
that optimal spring frequency is given by (for the relevant case, $n_\t{imp} \ll 1)$,
\begin{align}
	\Omega_\t{eff,opt} &\approx \omegam \left(\frac{n_\t{th}[\omegam]}{n_\t{imp}}\frac{1}{4 Q_0^2} \right)^{1/5} \\
	&= \omegam \left( \frac{\eta \nth}{Q_0} \right)^{1/5} \left( \frac{\omegasql}{\omegam} \right)^{2/5},
\end{align}
where, $Q_0 \equiv \omegam/\gammam[\omegam]$ is the intrinsic quality factor of the oscillator. 
Inserting this back into \cref{eq:neff:phase} gives,
\begin{align*}
	\neff +\frac{1}{2} &\approx \left( \frac{5}{2^{8/5}} (n_\t{th}^2 Q_0)^{2/5} \nimp^{1/5}
		+ \nba +\frac{1}{2} \right)\frac{\gammam[\omegaeff]}{\gammaeff}\\
		&\qquad + \nimp \frac{\gammaeff}{\gammam[\omegaeff]} \\
		&\geq 2\sqrt{\left( \frac{5}{2^{8/5}} (n_\t{th}^2 Q_0)^{2/5} \nimp^{1/5} 
		+ \nba +\frac{1}{2} \right) \nimp} \\
		&\geq 2\sqrt{ \frac{5}{2^{8/5}} (n_\t{th}^2 Q_0 \nimp^3)^{2/5} 
		+ \frac{1}{16}};
\end{align*}
here  the second line is the result of
optimizing over the feedback damping rate $\gammaeff$, while the last line is from the uncertainty principle 
(and we have omitted a small $\mathcal{O}(\nimp)$ term).
In order that $\neff < 1$, the last equation implies the requirement,
\begin{equation}
\label{eq:nimpQ}
	\nimp < \frac{\sqrt{2}}{5^{5/6}} \nth^{-2/3} Q_0^{-1/3},
\end{equation}
on the measurement sensitivity. For the experimentally relevant regime where the oscillator begins in a large thermal state,
i.e. $\nth \gg 1$, the requirement on the measurement sensitivity is slightly weaker for the case where
the oscillator is structurally damped and actively stiffened (scaling as $\nth^{-2/3}$) compared to the case of a velocity
damped oscillator (scaling as $\nth^{-1}$).
The condition for $\nimp$ in \cref{eq:nimpQ} can be rewritten as that for the mechanical Q-value,
\begin{equation}
    Q_0 > \left( \frac{5}{8} \right)^{5/4} \nth \left( \frac{\omegam}{\omegasql} \right)^3,
\end{equation}
or in terms of the oft-quoted ``$Qf$ product'', $Q_0 f_0 > (k_B T/h)\times (5/8)^{5/4} (\omegam/\omegasql)^3$. Note that
the necessary condition on the mechanical quality factor is relaxed for a low frequency oscillator
strongly coupled to a quantum-noise-limited optical field (i.e. $\omegasql \gg \omegam$). 
The unique $\omegasql^{-3}$ scaling on the 
Q-factor requirement is consistent with the idea that as 
$\omegasql$ increases, larger active spring frequencies can be realized; for a structurally damped oscillator, 
its thermal occupation reduces as $\omegaeff^{-1}$, while the penalty from feedback-back-action in realizing the spring
worsens as $\nimp \omegaeff^4 \propto (\omegaeff^2/\omegasql)^2$; 
their ratio is upper-bounded by a factor that scales as $\omegasql^{-3}$.

\subsection{General case: detuned readout with finite-bandwidth feedback}

The more general case harnesses the freedom to both detune the readout field from the optical cavity resonance --- which
produces an optical spring and rotates the quadrature of the the outgoing field with respect to the input field --- and
a variable-quadrature homodyne detection of the outgoing field --- which can be sensitive to the quantum correlations developed
via the radiation pressure interaction.
In this case, for a fixed optomechanical system, an experimenter has control 
over six parameters: 
the gain of the feedback filter $g_\t{fb}$ --- which effectively sets the feedback damping rate $\Gamma_\t{fb}$; 
the cutoff frequencies $\omegal$ and $\Omegah$ --- which together
with the feedback gain determines the feedback spring frequency $\Omega_\t{fb}$; 
the detuning, which contributes to the radiation pressure induced spring and damping; 
and the effective readout phase $\thetaref$. 

In the following we will interchangeably use the phonon number and the purity as figures of merit to assess the
quality of the quantum state that is realized.
The purity satisfies $0\leq \mu \leq 1$, where the upper (lower) bound corresponds to a maximally pure (mixed) state.
In the scenario we consider, where the initial state of the oscillator is Gaussian 
(specifically, assumed to be thermal), and measurement and feedback are linear in the oscillator's position, 
the state realized by feedback is also Gaussian. For Gaussian states, the purity is related to the average quantum number
of its thermal component as, $\mu^{-1}= 2n_\t{eff}+1$. Thus \cref{eq:neff} directly gives the inverse of the purity.
Note however than the conventionally employed criteria for having realized the ground state of motion, $n_\t{eff}< 1$,
corresponds to a purity of, $\mu > 1/3$. In the following we will employ purity as a figure of merit.

For fixed detuning, the dependence of the readout angle is through the imprecision and the imprecision-backaction correlations, 
\begin{align}\label{eq:angle_dep}
	n_\t{imp} &\equiv n_\t{imp}^{\theta_\t{eff}} = n_\t{imp}^{\pi/2} (1+\cot^2 \theta_\t{eff}) \\
	n_\t{cor} &\equiv n_\t{cor}^{\theta_\t{eff}} = n_\t{cor}^{\pi/4} \cot \theta_\t{eff},
\end{align}
where, $n_\t{imp}^{\pi/2} = 1/(16\eta n_\t{ba})$ is the imprecision for conventional phase quadrature readout, and 
$n_\t{cor}^{\pi/4} = 1/(2\sqrt{\eta}) $. Clearly, phase quadrature readout ($\thetaref = \pi/2$) minimizes imprecision 
without harnessing any quantum correlations ($n_\t{cor}^{\pi/2} = 0$), while amplitude quadrature readout contains no 
information about the motion (i.e. $n_\t{imp}^0 \rightarrow \infty$). 
%so that the large quantum correlations potentially observable near amplitude quadrature are in vain. 
In the context of displacement measurement, the tradeoff between these two
scenarios is the principle of so-called ``variational measurement'' that can realize displacement sensitivity better
than that by phase quadrature readout \cite{VyatZub95,KLMTV,KamReg17,SudKip17,MasSch19}. 
Improved displacement sensitivity, in the context of feedback control, produces less feedback back-action; thus, optimizing
the readout angle to harness quantum correlations can lead to better state purity (with other parameters fixed).

Inserting \cref{eq:angle_dep} in \cref{eq:neff}, the latter can be put into the form,
\begin{equation}\label{eq:Rmu}
\begin{split}
	R \mu^{-1} =& \frac{\Ctot}{g}\left( \ntheff +\nba +\tfrac{1}{2} \right)\\ 
	&+ g \Cimp \nimp^{\pi/2}(1+\cot^2 \theta)\\
	&- \Ccor \ncor^{\pi/4} \cot \theta, 
\end{split}
\end{equation}
where, $g \equiv \Gamma_0[\omegaeff]/\gammaeff$ is the factor by which the damping rate has increased due to feedback,
$R\equiv 1 - \Omegah/\omegal$, and $C_\t{tot,imp,cor}$ are the dimensionless pre-factors for each of the three terms
in \cref{eq:neff}:
\begin{equation}\label{eq:Cdefs}
\begin{split}
	\Ctot &\equiv \frac{2\omegaeff^2 + (\omegal - \Omegah) \gammaeff + 2\omegal^2}{\omegal^2}\\
	\Cimp &\equiv \frac{2\omegaeff^2 + (\omegal - \Omegah) \gammaeff + 2\Omegah^2}{\omegaeff^2} \\
	\Ccor &\equiv \frac{2\omegaeff^2 + (\omegal -\Omegah) \gammaeff + 2\omegal \Omegah}{\omegal \omegaeff} .
\end{split}
\end{equation}
which are themselves functions of $g$, $\Omega_\t{H,L}$. 
% Note that these constant are constrained, for example,
% \begin{equation}
% 	\frac{\Ccor^2}{2\Cimp} = \frac{[1+(\Ctot -2)(\omegal/\Omegah)^2]^2}{1+\tfrac{1}{2}(\Ctot -2)(\omegal/\Omegah)^2}.
% \end{equation}
In this sense, the final occupation
depends on five parameters: effective readout angle (which includes the detuning), the increase in damping due to feedback, 
quantified by $g$, and the filter cutoff frequencies $\Omega_\t{H,L}$; the filter DC gain ($g_\t{fb}$) and spring
frequency ($\omegaeff$) can be determined in terms of these.

The optimal readout angle is defined to be the one that maximizes the final state purity $\mu$. Completing the
square in the angle-dependent terms of \cref{eq:Rmu}:
\begin{equation}\label{eq:RmuTheta}
\begin{split}
	R \mu^{-1} =& \frac{\Ctot}{g} \left( \ntheff +\nba +\tfrac{1}{2} \right) + g \Cimp \nimp^{\pi/2} \\
	&+ g \Cimp \nimp^{\pi/2} \left[
		\left( \cot \theta - \frac{\Ccor \ncor^{\pi/4}}{2g \Cimp \nimp^{\pi/2}} \right)^2 \right. \\
	&\qquad\qquad\qquad - \left. \left(\frac{\Ccor \ncor^{\pi/4}}{2g \Cimp \nimp^{\pi/2}} \right)^2 \right]\\
	\geq& \frac{\Ctot}{g} \left( \ntheff +\nba +\tfrac{1}{2} 
		- \frac{(\Ccor \ncor^{\pi/4})^2}{4 \Ctot \Cimp \nimp^{\pi/2}} \right) \\
	&+ g \Cimp \nimp^{\pi/2},
\end{split}
\end{equation}
where the inequality is true for the choice, $\cot \theta = \Ccor \ncor/(2g \Cimp \nimp^{\pi/2})$, which
minimizes the expression in the first line, and dictates the optimal readout angle.

The negative term in the first parenthesis of the last inequality above represents the decrease in back-action due
to back-action cancellation in variable-quadrature readout. Indeed re-writing the back-action related part inside that
parenthesis in the form,
\begin{align*}
	\nba - \frac{(\Ccor \ncor^{\pi/4})^2}{4 \Ctot \Cimp \nimp^{\pi/2}} 
	&= \nba \left( 1- \frac{(\Ccor \ncor^{\pi/4})^2}{4\Ctot \Cimp}\cdot \frac{1}{\nba \nimp^{\pi/2}} \right)\\
	&\geq \nba\left( 1- \frac{\Ccor^2}{\Ctot \Cimp} \right),
\end{align*}
shows the ideal efficacy of back-action evasion with variable-quadrature readout. 
(Here the inequality is a result of the statements of the uncertainty principle, $\nba \nimp^{\pi/2} \geq 1/16$, 
and $\ncor^{\pi/4} \leq 1/2$.)
Ideally, all back-action is cancelled, corresponding to the condition, $\Ccor^2 = \Ctot \Cimp$; as it turns out,
this happens when  \footnote{
	In order to see how this works out, it is essential to observe that $C_\t{tot,imp,cor}$ are 
	constrained: $\Ccor^2/(2 \Cimp) = \frac{[1+(\Ctot -2)(\omegal/\Omegah)^2]^2}{1+\tfrac{1}{2}(\Ctot -2)(\omegal/\Omegah)^2}$.
	Then, $1-\Ccor^2/(\Ctot \Cimp) \propto \Ctot -2.$
} $\Ctot = 2$. From \cref{eq:Cdefs}, and the fact that $\Omegah < \omegaeff < \omegal$, it follows that $\omegal \gtrsim 
\sqrt{\Omegah \gammaeff/2}$. This further implies that, $\Cimp \approx 2$. 

Thus, in the ideal case where these conditions can be met, all back-action can be suppressed, and so,
\begin{equation}
\begin{split}
	\mu^{-1} &\geq \frac{2}{g}\left( \ntheff +\tfrac{1}{2} \right) + 2g \nimp^{\pi/2} \\
	&\geq 4\sqrt{\left( \ntheff +\tfrac{1}{2} \right) \nimp^{\pi/2}}, 
\end{split}
\end{equation}
indicating that the ground-state can be realized if, $\nimp^{\pi/2} \lesssim (3/4)^2 \ntheff^{-1} = (3/4)^2 n_\t{th}(\Omega_0/\omegaeff)^2$, 
for a readout angle $\theta \approx \tfrac{\pi}{4}$.

\begin{figure}[ht!]
	\centering
	\includegraphics[width=\columnwidth]{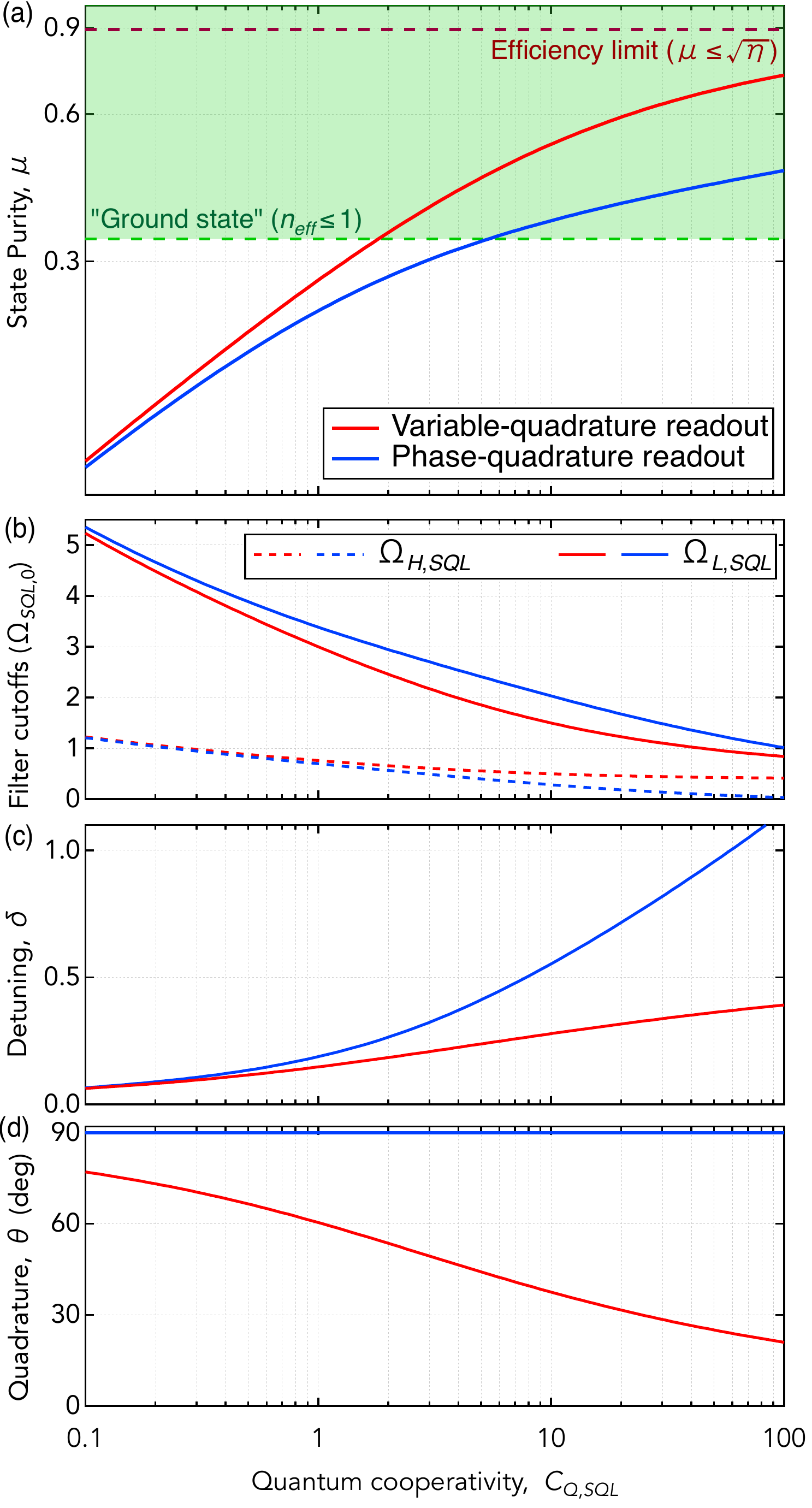}
	\caption{\label{fig:phononmin} 
	(a) Maximum achievable purity at any quantum cooperativity with active spring and detuned readout for a structurally
	damped oscillator. The red (blue) lines correspond to optimized (phase-fixed) readout angle in all panels.
	The region represents $n_\t{eff} < 1$ ($\mu < 1/3$). Brown dotted line indicates the limit to achievable purity
	due to inefficient detection, i.e. $\mu < \sqrt{\eta}$ ($\eta = 0.8$ here). 
	(b) shows the feedback filter cutoff frequencies in units of the SQL frequency $\omegasqlzero$; 
	solid lines are the low-pass cutoff
	and dahsed lines are the high-pass cutoff.
	(c) shows the optimal cavity detuning (normalized to FWHM).
	(d) shows the optimal readout angle.
	}
\end{figure}

The practical benefit of variable-quadrature readout is that for a given measurement imprecision, it can 
materialize moderate back-action cancellation so that the occupation achieved by feedback damping is 
lower than if phase readout were employed. To illustrate this
practical scenario, we numerically optimize the purity as a function of the five parameters: the two cutoff 
frequencies ($\omegal$ and $\Omegah$), the feedback gain ($\gammafb \simeq \gammaeff$), the normalized detuning ($\delta$), 
and the readout angle ($\thetaref$). 
The purity is optimized calculated for varying quantum cooperativities, $\Cqu \equiv \nba/\nth$. In order to emulate
conditions of fixed input power, the cutoff frequencies are normalized by the SQL frequency with zero detuning,
\begin{equation}
    \omegasqlzero = \sqrt{1 + \delta^2} \omegasql.
\end{equation}
For the same reason, we define $\Cqusql$ to be the quantum cooperativity at the SQL frequency with zero detuning. 
% This means that $\nba[\omegaeff]$ in \cref{eq:neff} is equal to $ \ntheff \Cqu[\omegaeff] 
% = \ntheff \frac{\Cqusql}{(1 + \delta^2)^2} \frac{\omegaeff}{\omegasqlzero}$.

% The cooperativity $\Cqu [\omegaeff]$ in \cref{eq:neff} is replaced with
% \begin{equation}
%     \Cqu [\omegaeff] = \frac{\Cqusql}{(1 + \delta^2)^2} \frac{\omegaeff}{\omegasqlzero}.
% \end{equation}
% The optical loss is fixed to be $\eta = 0.8$.

\Cref{fig:phononmin} shows the result of numerically optimizing the achievable purity as a function of 
quantum cooperativity $\Cqusql$. Green lines show the performance of phase quadrature readout ($\theta = \pi/2$), 
while blue shows the case where the readout laser is blue-detuned, and the cavity output is subjected to 
variable-quadrature homodyne measurement.
Variable-quadrature readout performs better in terms of the achievable state purity at all values of the cooperativity
(a result also know in the context of velocity-damped oscillators \cite{HabHam16}). 
Ground state cooling, where $\mu > 1/3$, can be achieved at $\Cqusql \gtrsim 1$, for a readout angle $\theta \approx \pi/3$.
The ultimate purity that can be achieved remains asymptotically bounded by $\mu < \sqrt{\eta}$.

At small (large) cooperativity, the optimal filter cutoff frequencies are relatively high (small), while the optimal
detuning is small (high). This implies that at small (large) cooperativity, the feedback (optical) spring must 
be dominant. The reason that is that in the small cooperativity regime, the optomechanical coupling 
is not strong enough to realize radiation pressure springs large enough to take advantage of the unique 
scaling of structural themal noise. Whereas in the high cooperativity regime, the feedback spring introduces additional 
decoherence from feedback back-action, so that optical spring is ideal in this regime.
In either case, the optimal spring frequency is around the SQL frequency.
% \Cref{fig:resratio} illustrates that this intuition is borne out. 
% %In this figure, we plot the effective resonant frequency normalized by the SQL frequency at the upper panel. 
% The optimal choice of spring frequency is around the SQL frequency.

% \begin{figure}
% 	\centering
% 	\includegraphics[width=\hsize]{fig_resratio.pdf}
% 	\caption{\label{fig:resratio}
% 	(a) The optimal value of the effective resonant frequencies (in units of the SQL frequency) with variable-quadrature 
% 	and phase readout. 
% 	(b) The ratios of the restoring force enhanced by the optical (dotted lines) and feedback (dashed lines) spring.
% 	\vs{have alphabetic subplot labels}}
% \end{figure}

% How that spring needs to be realized --- either by active feedback or cavity detuning --- is shown in the lower 
% panel of \cref{fig:resratio}. It depicts the ratio of the restoring forces acting on the oscillator due to optical
% and feedback springs, given respectively by, $\omegarp^2 / \omegaeff^2$ and $\Omegah \gammaeff / \omegaeff^2$. 
% In the small cooperativity regime, feedback spring should be dominant. That is because the optomechanical coupling 
% is not strong enough to realize radiation pressure springs large enough to take advantage of the unique 
% scaling of structural themal noise.
% In the high cooperativity regime, the feedback spring introduces additional decoherence from feedback back-action, 
% so that optical spring is ideal in this regime.

\section{Conclusion}

We have investigated the implications of structural damping on feedback-based motional ground-state preparation 
of elastically bound macroscopic mechanical oscillators. We find that the requirement to realize the ground
state is less stringent compared to the oft-studied case of a velocity-damped oscillator. 
That is because structural thermal noise reduces with increasing frequency much faster than velocity-proportional 
thermal noise. 
Hence actively stiffening the oscillator mode to take advantage of this decrease can be fruitful. 
However, that decrease comes at the expense of increasing back-action force fluctuations from the agency that realizes the
stiffened spring. The tradeoff between these competing sources of decoherence is optimized when the spring frequency 
is around the SQL frequency. Finally, feedback can be performed using a variable-quadrature homodyne
measurement of the oscillator's displacement, which outperforms feedback
based on phase-quadrature measurements at all values of the 
radiation-pressure cooperativity; this is due to back-action cancellation intrinsic to the variable-quadrature 
measurement scheme.

All of the above conclusions crucially rely on the implicit assumption that the favourable frequency-scaling of 
structural thermal noise continues well beyond the SQL frequency of the mechanical mode of interest. 
This necessitates careful suspension design to eliminate other mechanical modes in that vicinity.

These observations are directly relevant to experiments that hope to realize
pure quantum states of macroscopic mechanical oscillators to explore the interface between quantum physics and gravity.

\section{Acknowledgement}
We thank Hiroaki Ishizuka and Koji Nagano for fruitful discussions. KK is supported by JSPS Overseas Research Fellowship.

\appendix

\section{Imprecision-backaction product for arbitrary detuning}\label{sec:ImpBA}

In this section, we present the general form of the imprecision-backaction product for the displacement measurement of a mechanical oscillator at arbitrary detuning and homodyne angle. 
%and we further simplify it to the broadband cavity regime considered in this paper.

Let us consider a mechanical oscillator embedded as the end mirror of a single-sided optical cavity, pumped
by an ideal coherent state at the effective detuning $\Delta$. The intracavity optical fluctuations ($\delta a$) and
mechanical displacement fluctuations ($\delta x$) are described by the quantum Langevin equations \cite{Aspelmeyer2014}:
\begin{equation}\label{eq:eom_a}
    \delta\dot{{a}} = \left(i\Delta - \frac{\kappa}{2}\right)\delta{a} + \sqrt{\kappa}\delta{a}_\t{in} + iG\sqrt{\bar{n}}\delta{x},
\end{equation}
\begin{equation}\label{eq:eom_x}
    \delta\ddot{{x}} + \Gamma_\t{m}\delta\dot{{x}} + \Omega_\t{m}^2 \delta{x} = \frac{1}{m}(\delta{F}_\t{th} - \hbar G \sqrt{\bar{n}}(\delta{a}+\delta{a}^\dagger)).
\end{equation}
Note that \Cref{eq:eom_a} implies that the entry port is the dominant source of intracavity field losses. We may rewrite these equations in the Fourier domain as
\begin{equation}\label{eq:a_sol}
    \delta{a}[\Omega] = \frac{\sqrt{\kappa}\delta{a}_\t{in} + iG\sqrt{\bar{n}}\delta{x}}
    {-i(\Delta + \Omega) + \kappa/2},
\end{equation}
\begin{equation}\label{eq:x_sol}
    \delta{x}[\Omega] = \chi_\t{m} (\delta{F}_\t{th} + \delta{F}_\t{opt}).
\end{equation}
where $\chi_\t{m} = (m(\Omega_\t{m}^2 - \Omega^2 -i\Omega \Gamma_\t{m}))^{-1}$ is the intrinsic susceptibility of the mechanical oscillator, and $\delta{F}_\t{opt} = - \hbar G \sqrt{\bar{n}}(\delta{a}+\delta{a}^\dagger)$ is the total backaction 
force exerted on the mechanical oscillator due to the radiation pressure interaction. 
The reflected field, given by,
\begin{equation}\label{eq:bc_ref}
    \delta{a}_\t{ref} = \delta{a}_\t{in} - \sqrt{\kappa}\delta{a},
\end{equation}
is subjected to ideal homodyne detection with a local oscillator phase shifted by $\theta$; the resulting photocurrent
fluctuations are proportional to fluctuations of the quadrature,
\begin{equation}\label{eq:q_hom}
    \delta{q}^{\theta}_\t{ref} = \frac{1}{\sqrt{2}}\left(\delta{a}_\t{ref} e^{-i \theta} + \delta{a}^\dagger_\t{ref} e^{i \theta}\right).
\end{equation}

We may compute the spectral density of the homodyne quadrature as
\begin{equation}\label{eq:SD_def}
    \bar{S}^{\theta, \Delta, \t{ref}}_{qq} [\Omega] 2\pi \delta[0] = \langle \delta{q}^{\theta}_\t{ref}[\Omega] 
    \delta{q}^{\theta}_\t{ref}[-\Omega] \rangle,
\end{equation}
which may be written as 
\begin{equation}\label{eq:homodyne_spectrum}
    \begin{split}
        \bar{S}^{\theta, \Delta, \t{ref}}_{qq} [\Omega] \propto & \bar{S}^{\theta, \Delta, \t{imp}}_{xx} [\Omega] + \abs{\chi_\t{m}}^2 \left( \bar{S}^{\t{th}}_{xx} [\Omega] + \Bar{S}^{\delta, \t{rp}}_{FF}[\Omega] \right) \\
        & + 2\t{Re}\left( \chi_\t{m}\bar{S}^{\theta,\Delta,\t{rp,imp}}_{Fx} [\Omega] \right).
    \end{split}
\end{equation}
Computing this spectrum from \cref{eq:q_hom} following \cref{eq:SD_def} and noting the only non-zero correlator for the input vacuum fluctuations, $\langle \delta{a}_\t{in}[\Omega] \delta{a}^\dagger_\t{in}[-\Omega]\rangle = 2\pi\delta[0]$, we can identify the imprecision noise spectral density, $\bar{S}^{\theta, \Delta, \t{imp}}_{xx}$, the backaction force spectral density, $\Bar{S}^{\delta, \t{rp}}_{FF}$, as well as the correlation term, $\t{Re}\left(\bar{S}^{\theta,\Delta,\t{rp,imp}}_{Fx}\right)$ as follows.

Defining two frequency scale factors $\delta \equiv \frac{2\Delta}{\kappa}$ and $\omega \equiv \frac{2\Omega}{\kappa}$, the spectral density of the imprecision noise in \cref{eq:homodyne_spectrum} is given by
\begin{equation}
    \Bar{S}^{\theta, \delta, \t{imp}}_{xx}[\omega] = A^\t{imp} \Bar{S}^{\frac{\pi}{2}, 0, \t{imp}}_{xx}[\omega],
\end{equation}
where
\begin{equation}\label{eq:Aimp}
    A^\t{imp} = (1+\omega^2)^{-1} \frac{(1+(\delta + \omega)^2)(1+(\delta-\omega)^2)}{(\sin{\theta}-\delta\cos{\theta})^2 + 2\omega^2\sin^2(\theta)},
\end{equation}
\begin{equation}
    \Bar{S}^{\frac{\pi}{2}, 0, \t{imp}}_{xx}[\omega] = \left( \frac{\kappa}{16 G^2 \bar{n}} \right) (1+\omega^2).
\end{equation}
Next, the optical back-action force $\delta{F}_\t{rp}$ is the part of $\delta{F}_\t{opt}$ that only depends on the incoming vacuum field fluctuations, $\delta{a}_\t{in}$ and $\delta{a}^\dagger_\t{in}$.
%\begin{equation}\label{eq:Fba}
%    \delta{F}_\t{rp} = -\hbar G\sqrt{\bar{n}}\sqrt{\kappa}\left( \frac{\delta{a}_\t{in}}{-i(\Delta + \Omega) + \frac{\kappa}{2}} + \frac{\delta{a}^\dagger_\t{in}}{i(\Delta - \Omega) + \frac{\kappa}{2}}\right).
%\end{equation}
Its symmetrized spectral density as identified in \cref{eq:homodyne_spectrum} is expressed as
\begin{equation}
    \Bar{S}^{\delta, \t{rp}}_{FF}[\omega] = A^\t{rp} \Bar{S}^{0, \t{rp}}_{FF}[\omega],
\end{equation}
where 
\begin{equation}
    A^\t{rp} = (1+\omega^2)\frac{1+\delta^2+\omega^2}{(1+(\delta + \omega)^2)(1+(\delta-\omega)^2)},
\end{equation}
\begin{equation}
    \Bar{S}^{0, \t{rp}}_{FF}[\omega] = 4 \hbar \frac{G^2 \bar{n}}{\kappa}(1+\omega^2)^{-1}.
\end{equation}
Hence, the imprecision-backaction product in the general case of an arbitrary effective detuning and homodyne measurement angle 
is given by,
\begin{equation}\label{eq:full_SQL}
    \Bar{S}^{\theta, \delta, \t{imp}}_{xx}[\omega]\Bar{S}^{\delta, \t{rp}}_{FF}[\omega] = \frac{\hbar^2}{4} \frac{1+\delta^2 + \omega^2}{(\sin{\theta}-\delta\cos{\theta})^2 + 2\omega^2\sin^2(\theta)}.
\end{equation}
Note that the minimum value of the product is precisely $\hbar^2/4$, for $\delta = 0$ and $\theta = \pi/2)$; i.e.
resonant readout of the phase quadrature is ideal. 

In the broadband cavity regime (i.e. $\Omega \ll \kappa$ or $\omega \ll 1$), we have that,
\begin{equation}\label{eq:imp_bband}
        \Bar{S}^{\theta, \delta, \t{imp}}_{xx}[\omega] \approx \Bar{S}^{\frac{\pi}{2}, 0, \t{imp}}_{xx} \left( \frac{1+\delta^2}{\sin^2(\theta-\arctan{\delta})} + \mathcal{O}(\omega^2) \right),
\end{equation}
\begin{equation}\label{eq:ba_bband}
    \Bar{S}^{\delta, \t{rp}}_{FF}[\omega] \approx  \Bar{S}^{0, \t{rp}}_{FF} \left( \frac{1}{1+\delta^2} + \mathcal{O}(\omega^2) \right),
\end{equation}
and therefore,
\begin{equation}\label{eq:SQL_bband}
    \Bar{S}^{\theta, \delta, \t{imp}}_{xx}[\omega]\Bar{S}^{\delta, \t{rp}}_{FF}[\omega] \approx
    \frac{\hbar^2}{4}\csc^2(\theta-\arctan{\delta}),
\end{equation}
i.e. the effect of detuning, on the imprecision-backaction product in the broadband cavity regime, is equivalent to
a quadrature rotation $\arctan(2 \Delta/\kappa)$ by the cavity.
Furthermore, in the limit of $\delta, \omega \ll 1$, measuring the phase quadrature ($\theta = \pi/2$) is indeed the optimal strategy.

In the case of lossy homodyne detection, quantified by a non-unit detection efficiency $\eta \leq 1$, the
imprecision-backaction product is modified as: $[\bar{S}_{xx}^\t{imp} \bar{S}_\t{FF}^\t{rp}]_\eta = 
[\bar{S}_{xx}^\t{imp} \bar{S}_\t{FF}^\t{rp}]_{\eta=1}/\eta $. Thus, the effect of detuning
and general readout quadrature in \Cref{eq:SQL_bband} can be interpreted as an effective loss,
$\sin^2(\theta -\arctan \alpha)$.

Lastly, \cref{eq:homodyne_spectrum} allows us to identify the real part of the cross-correlation spectral density between the backaction force and the imprecision noise. 
%Using \cref{eq:ximp,eq:Fba}, we may calculate the unsymmetrized cross spectral density via
%\begin{equation}
%    S^{\theta,\Delta,\t{rp,imp}}_{Fx} [\Omega] 2\pi \delta[0] = \langle \delta{F}_\t{rp}[\Omega] 
%    \delta{x}_\t{imp}[-\Omega] \rangle.
%\end{equation}
%Symmetrizing,
%\begin{equation}
%    \Bar{S}^{\theta,\delta,\t{rp,imp}}_{Fx} [\omega] = \frac{1}{2} \left( S^{\theta,\delta,\t{rp,imp}}_{Fx} [\omega] + S^{\theta,\delta,\t{imp,rp}}_{xF} [-\omega] \right)
%        & = \frac{\hbar}{2} \Bigg( \frac{-(\cos{\theta} + \delta\sin{\theta})(\sin{\theta} - \delta\cos{\theta}) - \omega^2\sin{\theta}\cos{\theta}}{(\sin{\theta} - \delta\cos{\theta})^2 + 2 \omega^2\sin^2{\theta}} \\
%        & + \frac{i\delta\omega}{(\sin{\theta} - \delta\cos{\theta})^2 + 2 \omega^2\sin^2{\theta} } \Bigg).
%    \end{split}
%\end{equation}
In the broadband cavity regime ($\Omega \ll \kappa$ or $\omega \ll 1$), it is given by
\begin{equation}\label{eq:Fx}
    \t{Re}\left(\Bar{S}^{\theta,\delta,\t{rp,imp}}_{Fx} [\omega] \right) =  - \frac{\hbar}{2}  \left(  \cot(\theta - \arctan{\delta}) + \mathcal{O}(\omega^2) \right).
\end{equation}
Note that in the presence of lossy detection, the correlation term in \cref{eq:Fx} is modified as $[\bar{S}_\t{Fx}^\t{rp,imp}]_\eta = 
[\bar{S}_\t{Fx}^\t{rp,imp}]_{\eta=1}/\sqrt{\eta}$.

\section{Calculation of the mean phonon number}
\label{sec:AppA}

Here we describe the details of the integration of the power spectra $\bar{S}_{xx}$ of the oscillator under the combined
action of feedback and detuned optical spring.
Using the feedback filter in \cref{eq:chifb_new}, \cref{eq:dispspec_all} takes the form,
\begin{multline}
	\bar{S}_{xx}[\Omega] = \abs{\chi_\t{eff}[\Omega]}^2  \Bigg[ \left( 1 + \frac{1}{\Cqu [\omegaeff]} \right) \bar{S}_{FF}^\t{rp} \\
	+ \left( \frac{m\omegam^2 \omegal}{\Omegah} \right)^2 \frac{\Omega^2 + \Omegah^2}{\Omega^2 + \omegal^2} \gfb^2 \bar{S}_{xx}^\t{imp} + \\ \frac{2 m\omegam^2 \omegal}{\Omegah} \frac{\Omega^2 + \Omegah \omegal }{\Omega^2 + \omegal^2} \gfb \bar{S}_{Fx}^\t{rp,imp} \Bigg],
\end{multline}
where, $\Cqu [\omegaeff] = \bar{S}_{FF}^\t{rp}/S_{FF}^\t{th}[\omegaeff]$ is the quantum cooperativity. Considering that the typical feedback damping is dominant in the total effective damping ($\gammaeff \simeq \gammafb$), this can be recast as,
\begin{equation}
    \left( \Omega^2 + \omegal^2 \right) \abs{\chi_\t{eff}[\Omega]}^{-2} \bar{S}_{xx}[\Omega] = \Lambda_{\mathrm{L}} \Omega^2 + \Lambda_{\mathrm{H}} \omegal^2,
\end{equation}
where,
\begin{multline}
    \Lambda_{\mathrm{L}} = \left( 1 + \frac{1}{\Cqu [\omegaeff]} \right) \bar{S}_{FF}^\t{rp} + m^2 \omegal^2 \gammaeff^2 \bar{S}_{xx}^\t{imp} \\
    + 2 m \omegal \gammaeff \bar{S}_{Fx}^\t{rp,imp},
\end{multline}
\begin{multline}
    \Lambda_{\mathrm{H}} = \left( 1 + \frac{1}{\Cqu [\omegaeff]} \right) \bar{S}_{FF}^\t{rp} + m^2 \Omegah^2 \gammaeff^2 \bar{S}_{xx}^\t{imp} \\
    + 2 m \Omegah \gammaeff \bar{S}_{Fx}^\t{rp,imp}.
\end{multline}
The inverse of the effective susceptibility is represented as
\begin{multline}
    \frac{\omegal + i\Omega}{m} \chi_\t{eff}^{-1}[\Omega] = -i\Omega^3 - s_1 \Omega^2 + i s_2 \Omega + \omegal \omegaeff^2,
\end{multline}
where
\begin{gather}
    s_1 = \omegal + \gammam + \gammarp, \\
    s_2 = \omegaeff^2 + \left( \omegal - \Omegah \right) \gammaeff,
\end{gather}
and $\omegaeff^2 = \Omega_0^2 + \omegarp^2 + \Omegah \gammafb$.
In order to calculate the integration in \cref{eq:defxp}, we use the following identity \cite[3.112]{Gradshteyn1980},
\begin{equation}
\int^{\infty}_{-\infty} dx \frac{g_n (x)}{h_n(x) h_n(-x)} = (-1)^{n+1} \frac{\pi i}{a_0} \frac{M_n}{\Delta_n},
\end{equation}
where
\begin{gather}
g_n(x) = b_0 x^{2n-2} + b_1 x^{2n-4} + \cdots + b_{n-1}, \\
h_n(x) = a_0 x^n + a_1 x^{n-1} + \cdots + a_n, \\
\Delta_n = \left|
\begin{array}{ccccc}
a_1 & a_3 & a_5 & \cdots & 0 \\
a_0 & a_2 & a_4 & \ & 0 \\
0 & a_1 & a_3 & \ & 0 \\
\vdots & \ & \ & \ddots & \ \\
0 & 0 & 0 & \ & a_n
\end{array}
\right|, \\
M_n = \left|
\begin{array}{ccccc}
b_0 & b_1 & b_2 & \cdots & b_{n-1} \\
a_0 & a_2 & a_4 & \ & 0 \\
0 & a_1 & a_3 & \ & 0 \\
\vdots & \ & \ & \ddots & \ \\
0 & 0 & 0 & \ & a_n
\end{array}
\right|.
\end{gather}
Working to first order in $\Omega/\kappa$ (since we assume the system is in the broadband cavity regime), 
this identity can be applied with $n=3$. Here,
\begin{gather}
\Delta_3 = a_3 (a_1 a_2 - a_0 a_3), \\
M_3 = b_0 a_2 a_3 - b_1 a_0 a_3 + b_2 a_0 a_1.
\end{gather}
We use
\begin{align}
\begin{cases}
a_0 &= 1 \\
a_1 &= s_1 \\
a_2 &= s_2 \\
a_3 &= \omegal \omegaeff^2,
\end{cases}
\begin{cases}
b_0 &= 0 \\
b_1 &= \Lambda_{\mathrm{L}} \\
b_2 &= -\Lambda_{\mathrm{H}} \omegal^2.
\end{cases}
\end{align}
to calculate $\avg{\delta x^2}$, and
\begin{align}
\begin{cases}
a_0 &= 1 \\
a_1 &= s_1 \\
a_2 &= s_2 \\
a_3 &= \omegal \omegaeff^2,
\end{cases}
\begin{cases}
b_0 &= -\Lambda_{\mathrm{L}} \\
b_1 &= \Lambda_{\mathrm{H}} \omegal^2 \\
b_2 &= 0.
\end{cases}
\end{align}
to calculate $\avg{\delta p^2}$, respectively. The integrals are performed as
\begin{gather}
    \xsq = \frac{1}{2m^2 \omegaeff^2} \frac{\Lambda_{\mathrm{L}} \omegaeff^2 + \Lambda_{\mathrm{H}} \omegal s_1}{s_1 s_2 - \omegal \omegaeff^2}, \\
    \psq = \frac{1}{2} \frac{\Lambda_{\mathrm{L}} s_2 + \Lambda_{\mathrm{H}} \omegal^2}{s_1 s_2 - \omegal \omegaeff^2}.
\end{gather}
Typically we choose the low pass cutoff frequency of the filter which is much larger than the optical dissipation, so $s_1 \simeq \omegal$. Thus, the inverse of the purity is given by
\begin{equation}
    \mu^{-1} = \frac{\Lambda_{\mathrm{L}} \left( \omegaeff^2 + s_2 \right) + 2 \Lambda_{\mathrm{H}} \omegal^2}{2 \hbar m \omegaeff \omegal \left( s_2 - \omegaeff^2 \right)},
\end{equation}
which is the result in \cref{eq:neff}.

\bibliography{paper}

\begin{thebibliography}{50}%
\makeatletter
\providecommand \@ifxundefined [1]{%
 \@ifx{#1\undefined}
}%
\providecommand \@ifnum [1]{%
 \ifnum #1\expandafter \@firstoftwo
 \else \expandafter \@secondoftwo
 \fi
}%
\providecommand \@ifx [1]{%
 \ifx #1\expandafter \@firstoftwo
 \else \expandafter \@secondoftwo
 \fi
}%
\providecommand \natexlab [1]{#1}%
\providecommand \enquote  [1]{``#1''}%
\providecommand \bibnamefont  [1]{#1}%
\providecommand \bibfnamefont [1]{#1}%
\providecommand \citenamefont [1]{#1}%
\providecommand \href@noop [0]{\@secondoftwo}%
\providecommand \href [0]{\begingroup \@sanitize@url \@href}%
\providecommand \@href[1]{\@@startlink{#1}\@@href}%
\providecommand \@@href[1]{\endgroup#1\@@endlink}%
\providecommand \@sanitize@url [0]{\catcode `\\12\catcode `\$12\catcode
  `\&12\catcode `\#12\catcode `\^12\catcode `\_12\catcode `\%12\relax}%
\providecommand \@@startlink[1]{}%
\providecommand \@@endlink[0]{}%
\providecommand \url  [0]{\begingroup\@sanitize@url \@url }%
\providecommand \@url [1]{\endgroup\@href {#1}{\urlprefix }}%
\providecommand \urlprefix  [0]{URL }%
\providecommand \Eprint [0]{\href }%
\providecommand \doibase [0]{https://doi.org/}%
\providecommand \selectlanguage [0]{\@gobble}%
\providecommand \bibinfo  [0]{\@secondoftwo}%
\providecommand \bibfield  [0]{\@secondoftwo}%
\providecommand \translation [1]{[#1]}%
\providecommand \BibitemOpen [0]{}%
\providecommand \bibitemStop [0]{}%
\providecommand \bibitemNoStop [0]{.\EOS\space}%
\providecommand \EOS [0]{\spacefactor3000\relax}%
\providecommand \BibitemShut  [1]{\csname bibitem#1\endcsname}%
\let\auto@bib@innerbib\@empty
%</preamble>
\bibitem [{\citenamefont {Whittle}\ \emph {et~al.}(2021)\citenamefont
  {Whittle}, \citenamefont {Hall}, \citenamefont {Dwyer}, \citenamefont
  {Mavalvala}, \citenamefont {Sudhir},\ and\ \citenamefont {{LIGO Instrument
  Science Group}}}]{WhitSud21}%
  \BibitemOpen
  \bibfield  {author} {\bibinfo {author} {\bibfnamefont {C.}~\bibnamefont
  {Whittle}}, \bibinfo {author} {\bibfnamefont {E.~D.}\ \bibnamefont {Hall}},
  \bibinfo {author} {\bibfnamefont {S.}~\bibnamefont {Dwyer}}, \bibinfo
  {author} {\bibfnamefont {N.}~\bibnamefont {Mavalvala}}, \bibinfo {author}
  {\bibfnamefont {V.}~\bibnamefont {Sudhir}},\ and\ \bibinfo {author}
  {\bibnamefont {{LIGO Instrument Science Group}}},\ }\href
  {https://doi.org/10.1126/science.abh2634} {\bibfield  {journal} {\bibinfo
  {journal} {Science}\ }\textbf {\bibinfo {volume} {372}},\ \bibinfo {pages}
  {1333} (\bibinfo {year} {2021})}\BibitemShut {NoStop}%
\bibitem [{\citenamefont {Neuhaus}\ \emph {et~al.}(2021)\citenamefont
  {Neuhaus}, \citenamefont {Metzdorff}, \citenamefont {Zerkani}, \citenamefont
  {Chua}, \citenamefont {Teissier}, \citenamefont {Garcia-Sanchez},
  \citenamefont {Deleglise}, \citenamefont {Jacqmin}, \citenamefont {Briant},
  \citenamefont {Degallaix}, \citenamefont {Dolique}, \citenamefont {Cagnoli},
  \citenamefont {Traon}, \citenamefont {Chartier}, \citenamefont {Heidmann},\
  and\ \citenamefont {Cohadon}}]{Neu21}%
  \BibitemOpen
  \bibfield  {author} {\bibinfo {author} {\bibfnamefont {L.}~\bibnamefont
  {Neuhaus}}, \bibinfo {author} {\bibfnamefont {R.}~\bibnamefont {Metzdorff}},
  \bibinfo {author} {\bibfnamefont {S.}~\bibnamefont {Zerkani}}, \bibinfo
  {author} {\bibfnamefont {S.}~\bibnamefont {Chua}}, \bibinfo {author}
  {\bibfnamefont {J.}~\bibnamefont {Teissier}}, \bibinfo {author}
  {\bibfnamefont {D.}~\bibnamefont {Garcia-Sanchez}}, \bibinfo {author}
  {\bibfnamefont {S.}~\bibnamefont {Deleglise}}, \bibinfo {author}
  {\bibfnamefont {T.}~\bibnamefont {Jacqmin}}, \bibinfo {author} {\bibfnamefont
  {T.}~\bibnamefont {Briant}}, \bibinfo {author} {\bibfnamefont
  {J.}~\bibnamefont {Degallaix}}, \bibinfo {author} {\bibfnamefont
  {V.}~\bibnamefont {Dolique}}, \bibinfo {author} {\bibfnamefont
  {G.}~\bibnamefont {Cagnoli}}, \bibinfo {author} {\bibfnamefont {O.~L.}\
  \bibnamefont {Traon}}, \bibinfo {author} {\bibfnamefont {C.}~\bibnamefont
  {Chartier}}, \bibinfo {author} {\bibfnamefont {A.}~\bibnamefont {Heidmann}},\
  and\ \bibinfo {author} {\bibfnamefont {P.-F.}\ \bibnamefont {Cohadon}},\
  }\href {https://arxiv.org/abs/2104.11648v1} {\bibfield  {journal} {\bibinfo
  {journal} {arXiv:2104.11648}\ } (\bibinfo {year} {2021})}\BibitemShut
  {NoStop}%
\bibitem [{\citenamefont {Vinante}\ \emph {et~al.}(2008)\citenamefont
  {Vinante}, \citenamefont {Bignotto}, \citenamefont {Bonaldi}, \citenamefont
  {Cerdonio}, \citenamefont {Conti}, \citenamefont {Falferi}, \citenamefont
  {Liguori}, \citenamefont {Longo}, \citenamefont {Mezzena}, \citenamefont
  {Ortolan}, \citenamefont {Prodi}, \citenamefont {Salemi}, \citenamefont
  {Taffarello}, \citenamefont {Vedovato}, \citenamefont {Vitale},\ and\
  \citenamefont {Zendri}}]{auriga08}%
  \BibitemOpen
  \bibfield  {author} {\bibinfo {author} {\bibfnamefont {A.}~\bibnamefont
  {Vinante}}, \bibinfo {author} {\bibfnamefont {M.}~\bibnamefont {Bignotto}},
  \bibinfo {author} {\bibfnamefont {M.}~\bibnamefont {Bonaldi}}, \bibinfo
  {author} {\bibfnamefont {M.}~\bibnamefont {Cerdonio}}, \bibinfo {author}
  {\bibfnamefont {L.}~\bibnamefont {Conti}}, \bibinfo {author} {\bibfnamefont
  {P.}~\bibnamefont {Falferi}}, \bibinfo {author} {\bibfnamefont
  {N.}~\bibnamefont {Liguori}}, \bibinfo {author} {\bibfnamefont
  {S.}~\bibnamefont {Longo}}, \bibinfo {author} {\bibfnamefont
  {R.}~\bibnamefont {Mezzena}}, \bibinfo {author} {\bibfnamefont
  {A.}~\bibnamefont {Ortolan}}, \bibinfo {author} {\bibfnamefont {G.~A.}\
  \bibnamefont {Prodi}}, \bibinfo {author} {\bibfnamefont {F.}~\bibnamefont
  {Salemi}}, \bibinfo {author} {\bibfnamefont {L.}~\bibnamefont {Taffarello}},
  \bibinfo {author} {\bibfnamefont {G.}~\bibnamefont {Vedovato}}, \bibinfo
  {author} {\bibfnamefont {S.}~\bibnamefont {Vitale}},\ and\ \bibinfo {author}
  {\bibfnamefont {J.-P.}\ \bibnamefont {Zendri}},\ }\href
  {https://doi.org/10.1103/PhysRevLett.101.033601} {\bibfield  {journal}
  {\bibinfo  {journal} {Physical Review Letters}\ }\textbf {\bibinfo {volume}
  {101}},\ \bibinfo {pages} {033601} (\bibinfo {year} {2008})}\BibitemShut
  {NoStop}%
\bibitem [{\citenamefont {Diedrich}\ \emph {et~al.}(1989)\citenamefont
  {Diedrich}, \citenamefont {Bergquist}, \citenamefont {Itano},\ and\
  \citenamefont {Wineland}}]{Died89}%
  \BibitemOpen
  \bibfield  {author} {\bibinfo {author} {\bibfnamefont {F.}~\bibnamefont
  {Diedrich}}, \bibinfo {author} {\bibfnamefont {J.~C.}\ \bibnamefont
  {Bergquist}}, \bibinfo {author} {\bibfnamefont {W.~M.}\ \bibnamefont
  {Itano}},\ and\ \bibinfo {author} {\bibfnamefont {D.~J.}\ \bibnamefont
  {Wineland}},\ }\href {http://www.ncbi.nlm.nih.gov/pubmed/10040224} {\bibfield
   {journal} {\bibinfo  {journal} {Physical Review Letters}\ }\textbf {\bibinfo
  {volume} {62}},\ \bibinfo {pages} {403} (\bibinfo {year} {1989})}\BibitemShut
  {NoStop}%
\bibitem [{\citenamefont {Monroe}\ \emph {et~al.}(1995)\citenamefont {Monroe},
  \citenamefont {Meekhof}, \citenamefont {King}, \citenamefont {Jefferts},
  \citenamefont {Itano}, \citenamefont {Wineland},\ and\ \citenamefont
  {Gould}}]{Monroe95}%
  \BibitemOpen
  \bibfield  {author} {\bibinfo {author} {\bibfnamefont {C.}~\bibnamefont
  {Monroe}}, \bibinfo {author} {\bibfnamefont {D.}~\bibnamefont {Meekhof}},
  \bibinfo {author} {\bibfnamefont {B.}~\bibnamefont {King}}, \bibinfo {author}
  {\bibfnamefont {S.~R.}\ \bibnamefont {Jefferts}}, \bibinfo {author}
  {\bibfnamefont {W.~M.}\ \bibnamefont {Itano}}, \bibinfo {author}
  {\bibfnamefont {D.~J.}\ \bibnamefont {Wineland}},\ and\ \bibinfo {author}
  {\bibfnamefont {P.}~\bibnamefont {Gould}},\ }\href
  {http://www.atomwave.org/rmparticle/ao refs/aifm refs sorted by topic/atom
  trapping refs/Monroe95.pdf} {\bibfield  {journal} {\bibinfo  {journal}
  {Physical Review Letters}\ }\textbf {\bibinfo {volume} {75}},\ \bibinfo
  {pages} {4011} (\bibinfo {year} {1995})}\BibitemShut {NoStop}%
\bibitem [{\citenamefont {Perrin}\ \emph {et~al.}(1998)\citenamefont {Perrin},
  \citenamefont {Kuhn}, \citenamefont {Bouchoule},\ and\ \citenamefont
  {Salomon}}]{Perr98}%
  \BibitemOpen
  \bibfield  {author} {\bibinfo {author} {\bibfnamefont {H.}~\bibnamefont
  {Perrin}}, \bibinfo {author} {\bibfnamefont {A.}~\bibnamefont {Kuhn}},
  \bibinfo {author} {\bibfnamefont {I.}~\bibnamefont {Bouchoule}},\ and\
  \bibinfo {author} {\bibfnamefont {C.}~\bibnamefont {Salomon}},\ }\href
  {https://doi.org/10.1209/epl/i1998-00261-y} {\bibfield  {journal} {\bibinfo
  {journal} {Europhysics Letters}\ }\textbf {\bibinfo {volume} {42}},\ \bibinfo
  {pages} {395} (\bibinfo {year} {1998})}\BibitemShut {NoStop}%
\bibitem [{\citenamefont {Eschner}\ \emph {et~al.}(2003)\citenamefont
  {Eschner}, \citenamefont {Morigi}, \citenamefont {Schmidt-Kaler},\ and\
  \citenamefont {Blatt}}]{Esch03}%
  \BibitemOpen
  \bibfield  {author} {\bibinfo {author} {\bibfnamefont {J.}~\bibnamefont
  {Eschner}}, \bibinfo {author} {\bibfnamefont {G.}~\bibnamefont {Morigi}},
  \bibinfo {author} {\bibfnamefont {F.}~\bibnamefont {Schmidt-Kaler}},\ and\
  \bibinfo {author} {\bibfnamefont {R.}~\bibnamefont {Blatt}},\ }\href
  {https://doi.org/10.1364/JOSAB.20.001003} {\bibfield  {journal} {\bibinfo
  {journal} {JOSA B}\ }\textbf {\bibinfo {volume} {20}},\ \bibinfo {pages}
  {1003} (\bibinfo {year} {2003})}\BibitemShut {NoStop}%
\bibitem [{\citenamefont {Maunz}\ \emph {et~al.}(2004)\citenamefont {Maunz},
  \citenamefont {Puppe}, \citenamefont {Schuster}, \citenamefont {Syassen},
  \citenamefont {Pinkse},\ and\ \citenamefont {Rempe}}]{Maunz04}%
  \BibitemOpen
  \bibfield  {author} {\bibinfo {author} {\bibfnamefont {P.}~\bibnamefont
  {Maunz}}, \bibinfo {author} {\bibfnamefont {T.}~\bibnamefont {Puppe}},
  \bibinfo {author} {\bibfnamefont {I.}~\bibnamefont {Schuster}}, \bibinfo
  {author} {\bibfnamefont {N.}~\bibnamefont {Syassen}}, \bibinfo {author}
  {\bibfnamefont {P.~W.~H.}\ \bibnamefont {Pinkse}},\ and\ \bibinfo {author}
  {\bibfnamefont {G.}~\bibnamefont {Rempe}},\ }\href
  {http://www.nature.com/nature/journal/v428/n6978/abs/nature02387.html}
  {\bibfield  {journal} {\bibinfo  {journal} {Nature}\ }\textbf {\bibinfo
  {volume} {428}},\ \bibinfo {pages} {50} (\bibinfo {year} {2004})}\BibitemShut
  {NoStop}%
\bibitem [{\citenamefont {Boozer}\ \emph {et~al.}(2006)\citenamefont {Boozer},
  \citenamefont {Boca}, \citenamefont {Miller}, \citenamefont {Northup},\ and\
  \citenamefont {Kimble}}]{Booz06}%
  \BibitemOpen
  \bibfield  {author} {\bibinfo {author} {\bibfnamefont {A.~D.}\ \bibnamefont
  {Boozer}}, \bibinfo {author} {\bibfnamefont {A.}~\bibnamefont {Boca}},
  \bibinfo {author} {\bibfnamefont {R.}~\bibnamefont {Miller}}, \bibinfo
  {author} {\bibfnamefont {T.}~\bibnamefont {Northup}},\ and\ \bibinfo {author}
  {\bibfnamefont {H.~J.}\ \bibnamefont {Kimble}},\ }\href
  {https://doi.org/10.1103/PhysRevLett.97.083602} {\bibfield  {journal}
  {\bibinfo  {journal} {Physical Review Letters}\ }\textbf {\bibinfo {volume}
  {97}},\ \bibinfo {pages} {083602} (\bibinfo {year} {2006})}\BibitemShut
  {NoStop}%
\bibitem [{\citenamefont {Chan}\ \emph {et~al.}(2011)\citenamefont {Chan},
  \citenamefont {Alegre}, \citenamefont {Safavi-Naeini}, \citenamefont {Hill},
  \citenamefont {Krause}, \citenamefont {Gr{\"o}blacher}, \citenamefont
  {Aspelmeyer},\ and\ \citenamefont {Painter}}]{Chan2011}%
  \BibitemOpen
  \bibfield  {author} {\bibinfo {author} {\bibfnamefont {J.}~\bibnamefont
  {Chan}}, \bibinfo {author} {\bibfnamefont {T.~P.~M.}\ \bibnamefont {Alegre}},
  \bibinfo {author} {\bibfnamefont {A.~H.}\ \bibnamefont {Safavi-Naeini}},
  \bibinfo {author} {\bibfnamefont {J.~T.}\ \bibnamefont {Hill}}, \bibinfo
  {author} {\bibfnamefont {A.}~\bibnamefont {Krause}}, \bibinfo {author}
  {\bibfnamefont {S.}~\bibnamefont {Gr{\"o}blacher}}, \bibinfo {author}
  {\bibfnamefont {M.}~\bibnamefont {Aspelmeyer}},\ and\ \bibinfo {author}
  {\bibfnamefont {O.}~\bibnamefont {Painter}},\ }\href
  {https://doi.org/10.1038/nature10461} {\bibfield  {journal} {\bibinfo
  {journal} {Nature (London)}\ }\textbf {\bibinfo {volume} {478}},\ \bibinfo
  {pages} {89} (\bibinfo {year} {2011})}\BibitemShut {NoStop}%
\bibitem [{\citenamefont {Teufel}\ \emph {et~al.}(2011)\citenamefont {Teufel},
  \citenamefont {Donner}, \citenamefont {Li}, \citenamefont {Harlow},
  \citenamefont {Allman}, \citenamefont {Cicak}, \citenamefont {Sirois},
  \citenamefont {Whittaker}, \citenamefont {Lehnert},\ and\ \citenamefont
  {Simmonds}}]{Teufel2011}%
  \BibitemOpen
  \bibfield  {author} {\bibinfo {author} {\bibfnamefont {J.~D.}\ \bibnamefont
  {Teufel}}, \bibinfo {author} {\bibfnamefont {T.}~\bibnamefont {Donner}},
  \bibinfo {author} {\bibfnamefont {D.}~\bibnamefont {Li}}, \bibinfo {author}
  {\bibfnamefont {J.~W.}\ \bibnamefont {Harlow}}, \bibinfo {author}
  {\bibfnamefont {M.~S.}\ \bibnamefont {Allman}}, \bibinfo {author}
  {\bibfnamefont {K.}~\bibnamefont {Cicak}}, \bibinfo {author} {\bibfnamefont
  {A.~J.}\ \bibnamefont {Sirois}}, \bibinfo {author} {\bibfnamefont {J.~D.}\
  \bibnamefont {Whittaker}}, \bibinfo {author} {\bibfnamefont {K.~W.}\
  \bibnamefont {Lehnert}},\ and\ \bibinfo {author} {\bibfnamefont {R.~W.}\
  \bibnamefont {Simmonds}},\ }\href {https://doi.org/10.1038/nature10261}
  {\bibfield  {journal} {\bibinfo  {journal} {Nature (London)}\ }\textbf
  {\bibinfo {volume} {475}},\ \bibinfo {pages} {359} (\bibinfo {year}
  {2011})}\BibitemShut {NoStop}%
\bibitem [{\citenamefont {Kaufman}\ \emph {et~al.}(2012)\citenamefont
  {Kaufman}, \citenamefont {Lester},\ and\ \citenamefont {Regal}}]{Kauf12}%
  \BibitemOpen
  \bibfield  {author} {\bibinfo {author} {\bibfnamefont {A.~M.}\ \bibnamefont
  {Kaufman}}, \bibinfo {author} {\bibfnamefont {B.~J.}\ \bibnamefont
  {Lester}},\ and\ \bibinfo {author} {\bibfnamefont {C.}~\bibnamefont
  {Regal}},\ }\href {https://doi.org/10.1103/PhysRevX.2.041014} {\bibfield
  {journal} {\bibinfo  {journal} {Physical Review X}\ }\textbf {\bibinfo
  {volume} {2}},\ \bibinfo {pages} {041014} (\bibinfo {year}
  {2012})}\BibitemShut {NoStop}%
\bibitem [{\citenamefont {Peterson}\ \emph {et~al.}(2016)\citenamefont
  {Peterson}, \citenamefont {Purdy}, \citenamefont {Kampel}, \citenamefont
  {Andrews}, \citenamefont {Yu}, \citenamefont {Lehnert},\ and\ \citenamefont
  {Regal}}]{Peterson2016}%
  \BibitemOpen
  \bibfield  {author} {\bibinfo {author} {\bibfnamefont {R.~W.}\ \bibnamefont
  {Peterson}}, \bibinfo {author} {\bibfnamefont {T.~P.}\ \bibnamefont {Purdy}},
  \bibinfo {author} {\bibfnamefont {N.~S.}\ \bibnamefont {Kampel}}, \bibinfo
  {author} {\bibfnamefont {R.~W.}\ \bibnamefont {Andrews}}, \bibinfo {author}
  {\bibfnamefont {P.-L.}\ \bibnamefont {Yu}}, \bibinfo {author} {\bibfnamefont
  {K.~W.}\ \bibnamefont {Lehnert}},\ and\ \bibinfo {author} {\bibfnamefont
  {C.~A.}\ \bibnamefont {Regal}},\ }\href
  {https://doi.org/10.1103/PhysRevLett.116.063601} {\bibfield  {journal}
  {\bibinfo  {journal} {Phys. Rev. Lett.}\ }\textbf {\bibinfo {volume} {116}},\
  \bibinfo {pages} {063601} (\bibinfo {year} {2016})}\BibitemShut {NoStop}%
\bibitem [{\citenamefont {Rossi}\ \emph {et~al.}(2018)\citenamefont {Rossi},
  \citenamefont {Mason}, \citenamefont {Chen}, \citenamefont {Tsaturyan},\ and\
  \citenamefont {Schliesser}}]{Rossi2018}%
  \BibitemOpen
  \bibfield  {author} {\bibinfo {author} {\bibfnamefont {M.}~\bibnamefont
  {Rossi}}, \bibinfo {author} {\bibfnamefont {D.}~\bibnamefont {Mason}},
  \bibinfo {author} {\bibfnamefont {J.}~\bibnamefont {Chen}}, \bibinfo {author}
  {\bibfnamefont {Y.}~\bibnamefont {Tsaturyan}},\ and\ \bibinfo {author}
  {\bibfnamefont {A.}~\bibnamefont {Schliesser}},\ }\href
  {https://doi.org/10.1038/s41586-018-0643-8} {\bibfield  {journal} {\bibinfo
  {journal} {Nature (London)}\ }\textbf {\bibinfo {volume} {563}},\ \bibinfo
  {pages} {53} (\bibinfo {year} {2018})}\BibitemShut {NoStop}%
\bibitem [{\citenamefont {Deli{\'c}}\ \emph {et~al.}(2020)\citenamefont
  {Deli{\'c}}, \citenamefont {Reisenbauer}, \citenamefont {Dare}, \citenamefont
  {Grass}, \citenamefont {Vuleti{\'c}}, \citenamefont {Kiesel},\ and\
  \citenamefont {Aspelmeyer}}]{Delic20}%
  \BibitemOpen
  \bibfield  {author} {\bibinfo {author} {\bibfnamefont {U.}~\bibnamefont
  {Deli{\'c}}}, \bibinfo {author} {\bibfnamefont {M.}~\bibnamefont
  {Reisenbauer}}, \bibinfo {author} {\bibfnamefont {K.}~\bibnamefont {Dare}},
  \bibinfo {author} {\bibfnamefont {D.}~\bibnamefont {Grass}}, \bibinfo
  {author} {\bibfnamefont {V.}~\bibnamefont {Vuleti{\'c}}}, \bibinfo {author}
  {\bibfnamefont {N.}~\bibnamefont {Kiesel}},\ and\ \bibinfo {author}
  {\bibfnamefont {M.}~\bibnamefont {Aspelmeyer}},\ }\bibfield  {journal}
  {\bibinfo  {journal} {Science}\ }\href
  {https://doi.org/10.1126/science.aba3993} {10.1126/science.aba3993} (\bibinfo
  {year} {2020}),\ \bibinfo {note} {publisher: American Association for the
  Advancement of Science Section: Report}\BibitemShut {NoStop}%
\bibitem [{\citenamefont {Tebbenjohanns}\ \emph {et~al.}(2020)\citenamefont
  {Tebbenjohanns}, \citenamefont {Frimmer}, \citenamefont {Jain}, \citenamefont
  {Windey},\ and\ \citenamefont {Novotny}}]{Tebb20}%
  \BibitemOpen
  \bibfield  {author} {\bibinfo {author} {\bibfnamefont {F.}~\bibnamefont
  {Tebbenjohanns}}, \bibinfo {author} {\bibfnamefont {M.}~\bibnamefont
  {Frimmer}}, \bibinfo {author} {\bibfnamefont {V.}~\bibnamefont {Jain}},
  \bibinfo {author} {\bibfnamefont {D.}~\bibnamefont {Windey}},\ and\ \bibinfo
  {author} {\bibfnamefont {L.}~\bibnamefont {Novotny}},\ }\href
  {https://doi.org/10.1103/PhysRevLett.124.013603} {\bibfield  {journal}
  {\bibinfo  {journal} {Physical Review Letters}\ }\textbf {\bibinfo {volume}
  {124}},\ \bibinfo {pages} {013603} (\bibinfo {year} {2020})}\BibitemShut
  {NoStop}%
\bibitem [{\citenamefont {Magrini}\ \emph {et~al.}(2021)\citenamefont
  {Magrini}, \citenamefont {Rosenzweig}, \citenamefont {Bach}, \citenamefont
  {Deutschmann-Olek}, \citenamefont {Hofer}, \citenamefont {Hong},
  \citenamefont {Kiesel}, \citenamefont {Kugi},\ and\ \citenamefont
  {Aspelmeyer}}]{MagAsp21}%
  \BibitemOpen
  \bibfield  {author} {\bibinfo {author} {\bibfnamefont {L.}~\bibnamefont
  {Magrini}}, \bibinfo {author} {\bibfnamefont {P.}~\bibnamefont {Rosenzweig}},
  \bibinfo {author} {\bibfnamefont {C.}~\bibnamefont {Bach}}, \bibinfo {author}
  {\bibfnamefont {A.}~\bibnamefont {Deutschmann-Olek}}, \bibinfo {author}
  {\bibfnamefont {S.~G.}\ \bibnamefont {Hofer}}, \bibinfo {author}
  {\bibfnamefont {S.}~\bibnamefont {Hong}}, \bibinfo {author} {\bibfnamefont
  {N.}~\bibnamefont {Kiesel}}, \bibinfo {author} {\bibfnamefont
  {A.}~\bibnamefont {Kugi}},\ and\ \bibinfo {author} {\bibfnamefont
  {M.}~\bibnamefont {Aspelmeyer}},\ }\href
  {https://doi.org/10.1038/s41586-021-03602-3} {\bibfield  {journal} {\bibinfo
  {journal} {Nature}\ }\textbf {\bibinfo {volume} {595}},\ \bibinfo {pages}
  {373} (\bibinfo {year} {2021})}\BibitemShut {NoStop}%
\bibitem [{\citenamefont {Tebbenjohanns}\ \emph {et~al.}(2021)\citenamefont
  {Tebbenjohanns}, \citenamefont {Mattana}, \citenamefont {Rossi},
  \citenamefont {Frimmer},\ and\ \citenamefont {Novotny}}]{TebNov21}%
  \BibitemOpen
  \bibfield  {author} {\bibinfo {author} {\bibfnamefont {F.}~\bibnamefont
  {Tebbenjohanns}}, \bibinfo {author} {\bibfnamefont {M.~L.}\ \bibnamefont
  {Mattana}}, \bibinfo {author} {\bibfnamefont {M.}~\bibnamefont {Rossi}},
  \bibinfo {author} {\bibfnamefont {M.}~\bibnamefont {Frimmer}},\ and\ \bibinfo
  {author} {\bibfnamefont {L.}~\bibnamefont {Novotny}},\ }\href
  {https://doi.org/10.1038/s41586-021-03617-w} {\bibfield  {journal} {\bibinfo
  {journal} {Nature}\ }\textbf {\bibinfo {volume} {595}},\ \bibinfo {pages}
  {378} (\bibinfo {year} {2021})}\BibitemShut {NoStop}%
\bibitem [{\citenamefont {Millen}\ \emph {et~al.}(2014)\citenamefont {Millen},
  \citenamefont {Deesuwan}, \citenamefont {Barker},\ and\ \citenamefont
  {Anders}}]{Mill14}%
  \BibitemOpen
  \bibfield  {author} {\bibinfo {author} {\bibfnamefont {J.}~\bibnamefont
  {Millen}}, \bibinfo {author} {\bibfnamefont {T.}~\bibnamefont {Deesuwan}},
  \bibinfo {author} {\bibfnamefont {P.}~\bibnamefont {Barker}},\ and\ \bibinfo
  {author} {\bibfnamefont {J.}~\bibnamefont {Anders}},\ }\href
  {https://doi.org/10.1038/nnano.2014.82} {\bibfield  {journal} {\bibinfo
  {journal} {Nature Nano}\ }\textbf {\bibinfo {volume} {9}},\ \bibinfo {pages}
  {425} (\bibinfo {year} {2014})}\BibitemShut {NoStop}%
\bibitem [{\citenamefont {Fedorov}\ \emph {et~al.}(2018)\citenamefont
  {Fedorov}, \citenamefont {Sudhir}, \citenamefont {Schilling}, \citenamefont
  {Sch{\"u}tz}, \citenamefont {Wilson},\ and\ \citenamefont
  {Kippenberg}}]{Fed18}%
  \BibitemOpen
  \bibfield  {author} {\bibinfo {author} {\bibfnamefont {S.~A.}\ \bibnamefont
  {Fedorov}}, \bibinfo {author} {\bibfnamefont {V.}~\bibnamefont {Sudhir}},
  \bibinfo {author} {\bibfnamefont {R.}~\bibnamefont {Schilling}}, \bibinfo
  {author} {\bibfnamefont {H.}~\bibnamefont {Sch{\"u}tz}}, \bibinfo {author}
  {\bibfnamefont {D.~J.}\ \bibnamefont {Wilson}},\ and\ \bibinfo {author}
  {\bibfnamefont {T.~J.}\ \bibnamefont {Kippenberg}},\ }\href
  {https://doi.org/10.1016/j.physleta.2017.05.046} {\bibfield  {journal}
  {\bibinfo  {journal} {Physics Letters A}\ }\bibinfo {series} {Special {Issue}
  in memory of {Professor} {V}.{B}. {Braginsky}},\ \textbf {\bibinfo {volume}
  {382}},\ \bibinfo {pages} {2251} (\bibinfo {year} {2018})}\BibitemShut
  {NoStop}%
\bibitem [{\citenamefont {Mancini}\ \emph {et~al.}(1998)\citenamefont
  {Mancini}, \citenamefont {Vitali},\ and\ \citenamefont {Tombesi}}]{Manc98}%
  \BibitemOpen
  \bibfield  {author} {\bibinfo {author} {\bibfnamefont {S.}~\bibnamefont
  {Mancini}}, \bibinfo {author} {\bibfnamefont {D.}~\bibnamefont {Vitali}},\
  and\ \bibinfo {author} {\bibfnamefont {P.}~\bibnamefont {Tombesi}},\ }\href
  {http://journals.aps.org/prl/abstract/10.1103/PhysRevLett.80.688} {\bibfield
  {journal} {\bibinfo  {journal} {Physical Review Letters}\ }\textbf {\bibinfo
  {volume} {80}},\ \bibinfo {pages} {688} (\bibinfo {year} {1998})}\BibitemShut
  {NoStop}%
\bibitem [{\citenamefont {Marquardt}\ \emph {et~al.}(2007)\citenamefont
  {Marquardt}, \citenamefont {Chen}, \citenamefont {Clerk},\ and\ \citenamefont
  {Girvin}}]{Marq07}%
  \BibitemOpen
  \bibfield  {author} {\bibinfo {author} {\bibfnamefont {F.}~\bibnamefont
  {Marquardt}}, \bibinfo {author} {\bibfnamefont {J.}~\bibnamefont {Chen}},
  \bibinfo {author} {\bibfnamefont {A.}~\bibnamefont {Clerk}},\ and\ \bibinfo
  {author} {\bibfnamefont {S.}~\bibnamefont {Girvin}},\ }\href
  {https://doi.org/10.1103/PhysRevLett.99.093902} {\bibfield  {journal}
  {\bibinfo  {journal} {Physical Review Letters}\ }\textbf {\bibinfo {volume}
  {99}},\ \bibinfo {pages} {093902} (\bibinfo {year} {2007})}\BibitemShut
  {NoStop}%
\bibitem [{\citenamefont {Wilson-Rae}\ \emph {et~al.}(2007)\citenamefont
  {Wilson-Rae}, \citenamefont {Nooshi}, \citenamefont {Zwerger},\ and\
  \citenamefont {Kippenberg}}]{Wilson07}%
  \BibitemOpen
  \bibfield  {author} {\bibinfo {author} {\bibfnamefont {I.}~\bibnamefont
  {Wilson-Rae}}, \bibinfo {author} {\bibfnamefont {N.}~\bibnamefont {Nooshi}},
  \bibinfo {author} {\bibfnamefont {W.}~\bibnamefont {Zwerger}},\ and\ \bibinfo
  {author} {\bibfnamefont {T.}~\bibnamefont {Kippenberg}},\ }\href
  {https://doi.org/10.1103/PhysRevLett.99.093901} {\bibfield  {journal}
  {\bibinfo  {journal} {Physical Review Letters}\ }\textbf {\bibinfo {volume}
  {99}},\ \bibinfo {pages} {093901} (\bibinfo {year} {2007})}\BibitemShut
  {NoStop}%
\bibitem [{\citenamefont {Genes}\ \emph {et~al.}(2008)\citenamefont {Genes},
  \citenamefont {Vitali}, \citenamefont {Tombesi}, \citenamefont {Gigan},\ and\
  \citenamefont {Aspelmeyer}}]{Genes2008}%
  \BibitemOpen
  \bibfield  {author} {\bibinfo {author} {\bibfnamefont {C.}~\bibnamefont
  {Genes}}, \bibinfo {author} {\bibfnamefont {D.}~\bibnamefont {Vitali}},
  \bibinfo {author} {\bibfnamefont {P.}~\bibnamefont {Tombesi}}, \bibinfo
  {author} {\bibfnamefont {S.}~\bibnamefont {Gigan}},\ and\ \bibinfo {author}
  {\bibfnamefont {M.}~\bibnamefont {Aspelmeyer}},\ }\href
  {https://doi.org/10.1103/PhysRevA.77.033804} {\bibfield  {journal} {\bibinfo
  {journal} {Phys. Rev. A}\ }\textbf {\bibinfo {volume} {77}},\ \bibinfo
  {pages} {033804} (\bibinfo {year} {2008})}\BibitemShut {NoStop}%
\bibitem [{\citenamefont {Christian}(1966)}]{Christ66}%
  \BibitemOpen
  \bibfield  {author} {\bibinfo {author} {\bibfnamefont {R.}~\bibnamefont
  {Christian}},\ }\href {https://doi.org/10.1016/0042-207X(66)91162-6}
  {\bibfield  {journal} {\bibinfo  {journal} {Vacuum}\ }\textbf {\bibinfo
  {volume} {16}},\ \bibinfo {pages} {175} (\bibinfo {year} {1966})}\BibitemShut
  {NoStop}%
\bibitem [{\citenamefont {Beresnev}\ \emph {et~al.}(1990)\citenamefont
  {Beresnev}, \citenamefont {Chernyak},\ and\ \citenamefont
  {Fomyagin}}]{Beres90}%
  \BibitemOpen
  \bibfield  {author} {\bibinfo {author} {\bibfnamefont {S.~A.}\ \bibnamefont
  {Beresnev}}, \bibinfo {author} {\bibfnamefont {V.~G.}\ \bibnamefont
  {Chernyak}},\ and\ \bibinfo {author} {\bibfnamefont {G.~A.}\ \bibnamefont
  {Fomyagin}},\ }\href {https://doi.org/10.1017/S0022112090003007} {\bibfield
  {journal} {\bibinfo  {journal} {Journal of Fluid Mechanics}\ }\textbf
  {\bibinfo {volume} {219}},\ \bibinfo {pages} {405} (\bibinfo {year}
  {1990})}\BibitemShut {NoStop}%
\bibitem [{\citenamefont {Cavalleri}\ \emph {et~al.}(2010)\citenamefont
  {Cavalleri}, \citenamefont {Ciani}, \citenamefont {Dolesi}, \citenamefont
  {Hueller}, \citenamefont {Nicolodi}, \citenamefont {Tombolato}, \citenamefont
  {Vitale}, \citenamefont {Wass},\ and\ \citenamefont {Weber}}]{Cav10}%
  \BibitemOpen
  \bibfield  {author} {\bibinfo {author} {\bibfnamefont {A.}~\bibnamefont
  {Cavalleri}}, \bibinfo {author} {\bibfnamefont {G.}~\bibnamefont {Ciani}},
  \bibinfo {author} {\bibfnamefont {R.}~\bibnamefont {Dolesi}}, \bibinfo
  {author} {\bibfnamefont {M.}~\bibnamefont {Hueller}}, \bibinfo {author}
  {\bibfnamefont {D.}~\bibnamefont {Nicolodi}}, \bibinfo {author}
  {\bibfnamefont {D.}~\bibnamefont {Tombolato}}, \bibinfo {author}
  {\bibfnamefont {S.}~\bibnamefont {Vitale}}, \bibinfo {author} {\bibfnamefont
  {P.~J.}\ \bibnamefont {Wass}},\ and\ \bibinfo {author} {\bibfnamefont
  {W.~J.}\ \bibnamefont {Weber}},\ }\href
  {https://doi.org/10.1016/j.physleta.2010.06.041} {\bibfield  {journal}
  {\bibinfo  {journal} {Physics Letters A}\ }\textbf {\bibinfo {volume}
  {374}},\ \bibinfo {pages} {3365} (\bibinfo {year} {2010})}\BibitemShut
  {NoStop}%
\bibitem [{\citenamefont {Saulson}(1990)}]{Saul90}%
  \BibitemOpen
  \bibfield  {author} {\bibinfo {author} {\bibfnamefont {P.~R.}\ \bibnamefont
  {Saulson}},\ }\href
  {http://journals.aps.org/prd/abstract/10.1103/PhysRevD.42.2437} {\bibfield
  {journal} {\bibinfo  {journal} {Physical Review D}\ }\textbf {\bibinfo
  {volume} {42}},\ \bibinfo {pages} {2437} (\bibinfo {year}
  {1990})}\BibitemShut {NoStop}%
\bibitem [{\citenamefont {Cripe}\ \emph {et~al.}(2019)\citenamefont {Cripe},
  \citenamefont {Aggarwal}, \citenamefont {Lanza}, \citenamefont {Libson},
  \citenamefont {Singh}, \citenamefont {Heu}, \citenamefont {Follman},
  \citenamefont {Cole}, \citenamefont {Mavalvala},\ and\ \citenamefont
  {Corbitt}}]{Cripe2019}%
  \BibitemOpen
  \bibfield  {author} {\bibinfo {author} {\bibfnamefont {J.}~\bibnamefont
  {Cripe}}, \bibinfo {author} {\bibfnamefont {N.}~\bibnamefont {Aggarwal}},
  \bibinfo {author} {\bibfnamefont {R.}~\bibnamefont {Lanza}}, \bibinfo
  {author} {\bibfnamefont {A.}~\bibnamefont {Libson}}, \bibinfo {author}
  {\bibfnamefont {R.}~\bibnamefont {Singh}}, \bibinfo {author} {\bibfnamefont
  {P.}~\bibnamefont {Heu}}, \bibinfo {author} {\bibfnamefont {D.}~\bibnamefont
  {Follman}}, \bibinfo {author} {\bibfnamefont {G.~D.}\ \bibnamefont {Cole}},
  \bibinfo {author} {\bibfnamefont {N.}~\bibnamefont {Mavalvala}},\ and\
  \bibinfo {author} {\bibfnamefont {T.}~\bibnamefont {Corbitt}},\ }\href
  {https://doi.org/10.1038/s41586-019-1051-4} {\bibfield  {journal} {\bibinfo
  {journal} {Nature (London)}\ }\textbf {\bibinfo {volume} {568}},\ \bibinfo
  {pages} {364} (\bibinfo {year} {2019})}\BibitemShut {NoStop}%
\bibitem [{\citenamefont {Braginsky}\ \emph {et~al.}(1997)\citenamefont
  {Braginsky}, \citenamefont {Gorodetsky},\ and\ \citenamefont
  {Khalili}}]{BragKhal97}%
  \BibitemOpen
  \bibfield  {author} {\bibinfo {author} {\bibfnamefont {V.~B.}\ \bibnamefont
  {Braginsky}}, \bibinfo {author} {\bibfnamefont {M.~L.}\ \bibnamefont
  {Gorodetsky}},\ and\ \bibinfo {author} {\bibfnamefont {F.~Y.}\ \bibnamefont
  {Khalili}},\ }\href {https://doi.org/10.1016/S0375-9601(97)00413-1}
  {\bibfield  {journal} {\bibinfo  {journal} {Phys. Lett. A}\ }\textbf
  {\bibinfo {volume} {232}},\ \bibinfo {pages} {340} (\bibinfo {year}
  {1997})}\BibitemShut {NoStop}%
\bibitem [{\citenamefont {Corbitt}\ \emph {et~al.}(2007)\citenamefont
  {Corbitt}, \citenamefont {Wipf}, \citenamefont {Bodiya}, \citenamefont
  {Ottaway}, \citenamefont {Sigg}, \citenamefont {Smith}, \citenamefont
  {Whitcomb},\ and\ \citenamefont {Mavalvala}}]{Corbitt2007}%
  \BibitemOpen
  \bibfield  {author} {\bibinfo {author} {\bibfnamefont {T.}~\bibnamefont
  {Corbitt}}, \bibinfo {author} {\bibfnamefont {C.}~\bibnamefont {Wipf}},
  \bibinfo {author} {\bibfnamefont {T.}~\bibnamefont {Bodiya}}, \bibinfo
  {author} {\bibfnamefont {D.}~\bibnamefont {Ottaway}}, \bibinfo {author}
  {\bibfnamefont {D.}~\bibnamefont {Sigg}}, \bibinfo {author} {\bibfnamefont
  {N.}~\bibnamefont {Smith}}, \bibinfo {author} {\bibfnamefont
  {S.}~\bibnamefont {Whitcomb}},\ and\ \bibinfo {author} {\bibfnamefont
  {N.}~\bibnamefont {Mavalvala}},\ }\href
  {https://doi.org/10.1103/PhysRevLett.99.160801} {\bibfield  {journal}
  {\bibinfo  {journal} {Phys. Rev. Lett.}\ }\textbf {\bibinfo {volume} {99}},\
  \bibinfo {pages} {160801} (\bibinfo {year} {2007})}\BibitemShut {NoStop}%
\bibitem [{\citenamefont {Ni}\ \emph {et~al.}(2012)\citenamefont {Ni},
  \citenamefont {Norte}, \citenamefont {Wilson}, \citenamefont {Hood},
  \citenamefont {Chang}, \citenamefont {Painter},\ and\ \citenamefont
  {Kimble}}]{Ni12}%
  \BibitemOpen
  \bibfield  {author} {\bibinfo {author} {\bibfnamefont {K.-K.}\ \bibnamefont
  {Ni}}, \bibinfo {author} {\bibfnamefont {R.}~\bibnamefont {Norte}}, \bibinfo
  {author} {\bibfnamefont {D.~J.}\ \bibnamefont {Wilson}}, \bibinfo {author}
  {\bibfnamefont {J.~D.}\ \bibnamefont {Hood}}, \bibinfo {author}
  {\bibfnamefont {D.~E.}\ \bibnamefont {Chang}}, \bibinfo {author}
  {\bibfnamefont {O.}~\bibnamefont {Painter}},\ and\ \bibinfo {author}
  {\bibfnamefont {H.~J.}\ \bibnamefont {Kimble}},\ }\href
  {https://doi.org/10.1103/PhysRevLett.108.214302} {\bibfield  {journal}
  {\bibinfo  {journal} {Physical Review Letters}\ }\textbf {\bibinfo {volume}
  {108}},\ \bibinfo {pages} {214302} (\bibinfo {year} {2012})}\BibitemShut
  {NoStop}%
\bibitem [{\citenamefont {Wilson}\ \emph {et~al.}(2015)\citenamefont {Wilson},
  \citenamefont {Sudhir}, \citenamefont {Piro}, \citenamefont {Schilling},
  \citenamefont {Ghadimi},\ and\ \citenamefont {Kippenberg}}]{Wilson2015}%
  \BibitemOpen
  \bibfield  {author} {\bibinfo {author} {\bibfnamefont {D.~J.}\ \bibnamefont
  {Wilson}}, \bibinfo {author} {\bibfnamefont {V.}~\bibnamefont {Sudhir}},
  \bibinfo {author} {\bibfnamefont {N.}~\bibnamefont {Piro}}, \bibinfo {author}
  {\bibfnamefont {R.}~\bibnamefont {Schilling}}, \bibinfo {author}
  {\bibfnamefont {A.}~\bibnamefont {Ghadimi}},\ and\ \bibinfo {author}
  {\bibfnamefont {T.~J.}\ \bibnamefont {Kippenberg}},\ }\href
  {https://doi.org/10.1038/nature14672} {\bibfield  {journal} {\bibinfo
  {journal} {Nature (London)}\ }\textbf {\bibinfo {volume} {524}},\ \bibinfo
  {pages} {325} (\bibinfo {year} {2015})}\BibitemShut {NoStop}%
\bibitem [{\citenamefont {Braginsky}\ and\ \citenamefont
  {Vyatchanin}(2002)}]{BragVya02}%
  \BibitemOpen
  \bibfield  {author} {\bibinfo {author} {\bibfnamefont {V.~B.}\ \bibnamefont
  {Braginsky}}\ and\ \bibinfo {author} {\bibfnamefont {S.~P.}\ \bibnamefont
  {Vyatchanin}},\ }\href
  {https://www.sciencedirect.com/science/article/abs/pii/S0375960102000208}
  {\bibfield  {journal} {\bibinfo  {journal} {Phys. Lett. A}\ }\textbf
  {\bibinfo {volume} {293}},\ \bibinfo {pages} {228} (\bibinfo {year}
  {2002})}\BibitemShut {NoStop}%
\bibitem [{\citenamefont {Habibi}\ \emph {et~al.}(2016)\citenamefont {Habibi},
  \citenamefont {Zeuthen}, \citenamefont {Ghanaatshoar},\ and\ \citenamefont
  {Hammerer}}]{HabHam16}%
  \BibitemOpen
  \bibfield  {author} {\bibinfo {author} {\bibfnamefont {H.}~\bibnamefont
  {Habibi}}, \bibinfo {author} {\bibfnamefont {E.}~\bibnamefont {Zeuthen}},
  \bibinfo {author} {\bibfnamefont {M.}~\bibnamefont {Ghanaatshoar}},\ and\
  \bibinfo {author} {\bibfnamefont {K.}~\bibnamefont {Hammerer}},\ }\href
  {https://doi.org/10.1088/2040-8978/18/8/084004} {\bibfield  {journal}
  {\bibinfo  {journal} {Journal of Optics}\ }\textbf {\bibinfo {volume} {18}},\
  \bibinfo {pages} {084004} (\bibinfo {year} {2016})}\BibitemShut {NoStop}%
\bibitem [{\citenamefont {Aspelmeyer}\ \emph {et~al.}(2014)\citenamefont
  {Aspelmeyer}, \citenamefont {Kippenberg},\ and\ \citenamefont
  {Marquardt}}]{Aspelmeyer2014}%
  \BibitemOpen
  \bibfield  {author} {\bibinfo {author} {\bibfnamefont {M.}~\bibnamefont
  {Aspelmeyer}}, \bibinfo {author} {\bibfnamefont {T.~J.}\ \bibnamefont
  {Kippenberg}},\ and\ \bibinfo {author} {\bibfnamefont {F.}~\bibnamefont
  {Marquardt}},\ }\href {https://doi.org/10.1103/RevModPhys.86.1391} {\bibfield
   {journal} {\bibinfo  {journal} {Rev. Mod. Phys.}\ }\textbf {\bibinfo
  {volume} {86}},\ \bibinfo {pages} {1391} (\bibinfo {year}
  {2014})}\BibitemShut {NoStop}%
\bibitem [{\citenamefont {Botter}\ \emph {et~al.}(2012)\citenamefont {Botter},
  \citenamefont {Brooks}, \citenamefont {Brahms}, \citenamefont {Schreppler},\
  and\ \citenamefont {Stamper-Kurn}}]{BottStam12}%
  \BibitemOpen
  \bibfield  {author} {\bibinfo {author} {\bibfnamefont {T.}~\bibnamefont
  {Botter}}, \bibinfo {author} {\bibfnamefont {D.~W.~C.}\ \bibnamefont
  {Brooks}}, \bibinfo {author} {\bibfnamefont {N.}~\bibnamefont {Brahms}},
  \bibinfo {author} {\bibfnamefont {S.}~\bibnamefont {Schreppler}},\ and\
  \bibinfo {author} {\bibfnamefont {D.~M.}\ \bibnamefont {Stamper-Kurn}},\
  }\href {https://doi.org/10.1103/PhysRevA.85.013812} {\bibfield  {journal}
  {\bibinfo  {journal} {Physical Review A}\ }\textbf {\bibinfo {volume} {85}},\
  \bibinfo {pages} {013812} (\bibinfo {year} {2012})}\BibitemShut {NoStop}%
\bibitem [{\citenamefont {Braginsky}\ and\ \citenamefont
  {Khalili}(1992)}]{brag}%
  \BibitemOpen
  \bibfield  {author} {\bibinfo {author} {\bibfnamefont {V.~B.}\ \bibnamefont
  {Braginsky}}\ and\ \bibinfo {author} {\bibfnamefont {F.~Y.}\ \bibnamefont
  {Khalili}},\ }\href@noop {} {\emph {\bibinfo {title} {Quantum
  Measurement}}},\ edited by\ \bibinfo {editor} {\bibfnamefont {K.~S.}\
  \bibnamefont {Thorne}}\ (\bibinfo  {publisher} {Cambridge University Press},\
  \bibinfo {year} {1992})\BibitemShut {NoStop}%
\bibitem [{\citenamefont {Clerk}\ \emph {et~al.}(2010)\citenamefont {Clerk},
  \citenamefont {Devoret}, \citenamefont {Girvin}, \citenamefont {Marquardt},\
  and\ \citenamefont {Schoelkopf}}]{clerk10}%
  \BibitemOpen
  \bibfield  {author} {\bibinfo {author} {\bibfnamefont {A.~A.}\ \bibnamefont
  {Clerk}}, \bibinfo {author} {\bibfnamefont {M.~H.}\ \bibnamefont {Devoret}},
  \bibinfo {author} {\bibfnamefont {S.~M.}\ \bibnamefont {Girvin}}, \bibinfo
  {author} {\bibfnamefont {F.}~\bibnamefont {Marquardt}},\ and\ \bibinfo
  {author} {\bibfnamefont {R.~J.}\ \bibnamefont {Schoelkopf}},\ }\href
  {https://doi.org/10.1103/RevModPhys.82.1155} {\bibfield  {journal} {\bibinfo
  {journal} {Reviews of Modern Physics}\ }\textbf {\bibinfo {volume} {82}},\
  \bibinfo {pages} {1155} (\bibinfo {year} {2010})}\BibitemShut {NoStop}%
\bibitem [{\citenamefont {Du~Pr{\'e}}(1950)}]{Pre50}%
  \BibitemOpen
  \bibfield  {author} {\bibinfo {author} {\bibfnamefont {F.~K.}\ \bibnamefont
  {Du~Pr{\'e}}},\ }\href {https://doi.org/10.1103/PhysRev.78.615} {\bibfield
  {journal} {\bibinfo  {journal} {Physical Review}\ }\textbf {\bibinfo {volume}
  {78}},\ \bibinfo {pages} {615} (\bibinfo {year} {1950})}\BibitemShut
  {NoStop}%
\bibitem [{\citenamefont {Mandelbrot}\ and\ \citenamefont
  {Ness}(1968)}]{mandelbrot_fractional_1968}%
  \BibitemOpen
  \bibfield  {author} {\bibinfo {author} {\bibfnamefont {B.}~\bibnamefont
  {Mandelbrot}}\ and\ \bibinfo {author} {\bibfnamefont {J.~V.}\ \bibnamefont
  {Ness}},\ }\href {http://epubs.siam.org/doi/pdf/10.1137/1010093} {\bibfield
  {journal} {\bibinfo  {journal} {SIAM Review}\ }\textbf {\bibinfo {volume}
  {10}},\ \bibinfo {pages} {422} (\bibinfo {year} {1968})}\BibitemShut
  {NoStop}%
\bibitem [{\citenamefont {Nelkin}\ and\ \citenamefont
  {Tremblay}(1981)}]{nelkin_deviation_1981}%
  \BibitemOpen
  \bibfield  {author} {\bibinfo {author} {\bibfnamefont {M.}~\bibnamefont
  {Nelkin}}\ and\ \bibinfo {author} {\bibfnamefont {A.~M.~S.}\ \bibnamefont
  {Tremblay}},\ }\href {https://doi.org/10.1007/BF01022186} {\bibfield
  {journal} {\bibinfo  {journal} {Journal of Statistical Physics}\ }\textbf
  {\bibinfo {volume} {25}},\ \bibinfo {pages} {253} (\bibinfo {year}
  {1981})}\BibitemShut {NoStop}%
\bibitem [{\citenamefont {Vitali}\ \emph {et~al.}(2003)\citenamefont {Vitali},
  \citenamefont {Mancini}, \citenamefont {Ribichini},\ and\ \citenamefont
  {Tombesi}}]{VitTom03}%
  \BibitemOpen
  \bibfield  {author} {\bibinfo {author} {\bibfnamefont {D.}~\bibnamefont
  {Vitali}}, \bibinfo {author} {\bibfnamefont {S.}~\bibnamefont {Mancini}},
  \bibinfo {author} {\bibfnamefont {L.}~\bibnamefont {Ribichini}},\ and\
  \bibinfo {author} {\bibfnamefont {P.}~\bibnamefont {Tombesi}},\ }\href
  {https://doi.org/10.1364/JOSAB.20.001054} {\bibfield  {journal} {\bibinfo
  {journal} {Journal of the Optical Society of America B}\ }\textbf {\bibinfo
  {volume} {20}},\ \bibinfo {pages} {1054} (\bibinfo {year}
  {2003})}\BibitemShut {NoStop}%
\bibitem [{\citenamefont {Sudhir}\ \emph {et~al.}(2017)\citenamefont {Sudhir},
  \citenamefont {Schilling}, \citenamefont {Fedorov}, \citenamefont
  {Sch{\"u}tz}, \citenamefont {Wilson},\ and\ \citenamefont
  {Kippenberg}}]{SudKip17}%
  \BibitemOpen
  \bibfield  {author} {\bibinfo {author} {\bibfnamefont {V.}~\bibnamefont
  {Sudhir}}, \bibinfo {author} {\bibfnamefont {R.}~\bibnamefont {Schilling}},
  \bibinfo {author} {\bibfnamefont {S.}~\bibnamefont {Fedorov}}, \bibinfo
  {author} {\bibfnamefont {H.}~\bibnamefont {Sch{\"u}tz}}, \bibinfo {author}
  {\bibfnamefont {D.}~\bibnamefont {Wilson}},\ and\ \bibinfo {author}
  {\bibfnamefont {T.}~\bibnamefont {Kippenberg}},\ }\href
  {https://doi.org/10.1103/PhysRevX.7.031055} {\bibfield  {journal} {\bibinfo
  {journal} {Physical Review X}\ }\textbf {\bibinfo {volume} {7}},\ \bibinfo
  {pages} {031055} (\bibinfo {year} {2017})}\BibitemShut {NoStop}%
\bibitem [{\citenamefont {Kampel}\ \emph {et~al.}(2017)\citenamefont {Kampel},
  \citenamefont {Peterson}, \citenamefont {Fischer}, \citenamefont {Yu},
  \citenamefont {Cicak}, \citenamefont {Simmonds}, \citenamefont {Lehnert},\
  and\ \citenamefont {Regal}}]{KamReg17}%
  \BibitemOpen
  \bibfield  {author} {\bibinfo {author} {\bibfnamefont {N.}~\bibnamefont
  {Kampel}}, \bibinfo {author} {\bibfnamefont {R.}~\bibnamefont {Peterson}},
  \bibinfo {author} {\bibfnamefont {R.}~\bibnamefont {Fischer}}, \bibinfo
  {author} {\bibfnamefont {P.-L.}\ \bibnamefont {Yu}}, \bibinfo {author}
  {\bibfnamefont {K.}~\bibnamefont {Cicak}}, \bibinfo {author} {\bibfnamefont
  {R.}~\bibnamefont {Simmonds}}, \bibinfo {author} {\bibfnamefont
  {K.}~\bibnamefont {Lehnert}},\ and\ \bibinfo {author} {\bibfnamefont
  {C.}~\bibnamefont {Regal}},\ }\href
  {https://doi.org/10.1103/PhysRevX.7.021008} {\bibfield  {journal} {\bibinfo
  {journal} {Physical Review X}\ }\textbf {\bibinfo {volume} {7}},\ \bibinfo
  {pages} {021008} (\bibinfo {year} {2017})}\BibitemShut {NoStop}%
\bibitem [{\citenamefont {Mason}\ \emph {et~al.}(2019)\citenamefont {Mason},
  \citenamefont {Chen}, \citenamefont {Rossi}, \citenamefont {Tsaturyan},\ and\
  \citenamefont {Schliesser}}]{MasSch19}%
  \BibitemOpen
  \bibfield  {author} {\bibinfo {author} {\bibfnamefont {D.}~\bibnamefont
  {Mason}}, \bibinfo {author} {\bibfnamefont {J.}~\bibnamefont {Chen}},
  \bibinfo {author} {\bibfnamefont {M.}~\bibnamefont {Rossi}}, \bibinfo
  {author} {\bibfnamefont {Y.}~\bibnamefont {Tsaturyan}},\ and\ \bibinfo
  {author} {\bibfnamefont {A.}~\bibnamefont {Schliesser}},\ }\href
  {https://doi.org/10.1038/s41567-019-0533-5} {\bibfield  {journal} {\bibinfo
  {journal} {Nature Physics}\ }\textbf {\bibinfo {volume} {15}},\ \bibinfo
  {pages} {745} (\bibinfo {year} {2019})}\BibitemShut {NoStop}%
\bibitem [{\citenamefont {Vyatchanin}\ and\ \citenamefont
  {Zubova}(1995)}]{VyatZub95}%
  \BibitemOpen
  \bibfield  {author} {\bibinfo {author} {\bibfnamefont {S.~P.}\ \bibnamefont
  {Vyatchanin}}\ and\ \bibinfo {author} {\bibfnamefont {E.~A.}\ \bibnamefont
  {Zubova}},\ }\href {https://doi.org/10.1016/0375-9601(95)00280-G} {\bibfield
  {journal} {\bibinfo  {journal} {Physics Letters A}\ }\textbf {\bibinfo
  {volume} {201}},\ \bibinfo {pages} {269} (\bibinfo {year}
  {1995})}\BibitemShut {NoStop}%
\bibitem [{\citenamefont {Kimble}\ \emph {et~al.}(2001)\citenamefont {Kimble},
  \citenamefont {Levin}, \citenamefont {Matsko}, \citenamefont {Thorne},\ and\
  \citenamefont {Vyatchanin}}]{KLMTV}%
  \BibitemOpen
  \bibfield  {author} {\bibinfo {author} {\bibfnamefont {H.~J.}\ \bibnamefont
  {Kimble}}, \bibinfo {author} {\bibfnamefont {Y.}~\bibnamefont {Levin}},
  \bibinfo {author} {\bibfnamefont {A.~B.}\ \bibnamefont {Matsko}}, \bibinfo
  {author} {\bibfnamefont {K.~S.}\ \bibnamefont {Thorne}},\ and\ \bibinfo
  {author} {\bibfnamefont {S.~P.}\ \bibnamefont {Vyatchanin}},\ }\href
  {https://doi.org/10.1103/PhysRevD.65.022002} {\bibfield  {journal} {\bibinfo
  {journal} {Physical Review D}\ }\textbf {\bibinfo {volume} {65}},\ \bibinfo
  {pages} {022002} (\bibinfo {year} {2001})}\BibitemShut {NoStop}%
\bibitem [{Note1()}]{Note1}%
  \BibitemOpen
  \bibinfo {note} {In order to see how this works out, it is essential to
  observe that $C_\protect \text {{tot,imp,cor}}$ are constrained: $C_\protect
  \mathrm {cor}^2/(2 C_\protect \mathrm {imp}) = \protect \frac {[1+(C_\protect
  \mathrm {tot}-2)(\Omega _{\protect \mathrm {L}}/\Omega _{\protect \mathrm
  {H}})^2]^2}{1+\protect \genfrac {}{}{}1{1}{2}(C_\protect \mathrm
  {tot}-2)(\Omega _{\protect \mathrm {L}}/\Omega _{\protect \mathrm {H}})^2}$.
  Then, $1-C_\protect \mathrm {cor}^2/(C_\protect \mathrm {tot}C_\protect
  \mathrm {imp}) \propto C_\protect \mathrm {tot}-2.$}\BibitemShut {NoStop}%
\bibitem [{\citenamefont {Gradshteyn}\ and\ \citenamefont
  {Ryzhik}(1980)}]{Gradshteyn1980}%
  \BibitemOpen
  \bibfield  {author} {\bibinfo {author} {\bibfnamefont {I.~S.}\ \bibnamefont
  {Gradshteyn}}\ and\ \bibinfo {author} {\bibfnamefont {I.~M.}\ \bibnamefont
  {Ryzhik}},\ }\href@noop {} {\bibfield  {journal} {\bibinfo  {journal} {{{\it
  Table of Integrals, Series and Products}, Academic Press, Orland}}\ ,\
  \bibinfo {pages} {p253}} (\bibinfo {year} {1980})}\BibitemShut {NoStop}%
\end{thebibliography}%
\bibliographystyle{apsrev.bst}
\end{document}